\documentclass[twocolumn,preprintnumbers,amsmath,amssymb]{revtex4}  

\usepackage{graphicx}
\usepackage{dcolumn}
\usepackage{bm}

\begin{document}

\title{Nematic superconductivity in Cu$_{x}$Bi$_{2}$Se$_{3}$: The surface Andreev bound states}

\author{Lei Hao$^{1,2}$ and C. S. Ting$^{1}$}
 \address{$^1$Department of Physics and Texas Center for Superconductivity, University of Houston, Houston, Texas 77204, USA \notag \\
 $^2$Department of Physics, Southeast University, Nanjing 210096, China \notag}

\date{\today}%

\begin{abstract}
We study theoretically the topological surface states (TSSs) and the possible surface Andreev bound states (SABSs) of Cu$_{x}$Bi$_{2}$Se$_{3}$ which is known to be a topological insulator at $x=0$. The superconductivity (SC) pairing of this compound is assumed to have the broken spin-rotation symmetry, similar to that of the A-phase of $^{3}$He as suggested by recent nuclear-magnetic resonance experiments. For both spheroidal and corrugated cylindrical Fermi surfaces with the hexagonal warping terms, we show that the bulk SC gap is rather anisotropic; the minimum of the gap is negligibly small as comparing to the maximum of the gap. This would make the fully-gapped pairing effectively nodal. For a clean system, our results indicate the bulk of this compound to be a topological superconductor with the SABSs appearing inside the bulk SC gap. The zero-energy SABSs which are Majorana fermions, together with the TSSs not gapped by the pairing, produce a zero-energy peak in the surface density of states (SDOS). The SABSs are expected to be stable against short-range nonmagnetic impurities, and the local SDOS is calculated around a nonmagnetic impurity. The relevance of our results to experiments is discussed.
\end{abstract}

\pacs{}

\maketitle

\section{\label{introduction}introduction}

Cu$_{x}$Bi$_{2}$Se$_{3}$, the first superconductor (SC) realized in a doped topological insulator, has attracted great interest since its discovery \cite{hor10,wray10}. After early gathering of experimental and theoretical evidences trying to imply the superconductivity (SC) pairing in this compound to be topologically nontrivial \cite{sasaki11,koren11,kirzhner12,kriener11,das11,bay12}, several later experiments seem to conclude that the actual pairing symmetry is the conventional $s$-wave and topologically trivial \cite{levy12,peng13}. However, initiated by a recent nuclear magnetic resonance (NMR) experiment \cite{matano16,fu16}, a new surge of researches on this material and its several variants (e.g., Sr$_{x}$Bi$_{2}$Se$_{3}$ \cite{zhao14,pan16,du17}, Nb$_{x}$Bi$_{2}$Se$_{3}$ \cite{qiu15,asaba17,lawson16}, and Tl$_{x}$Bi$_{2}$Se$_{3}$ \cite{wang16}) revived the possibility that the SC  phase realized in this group of materials is topologically nontrivial.
In the NMR experiment, a prominent in-plane uniaxial anisotropy in the SC order parameter is observed in the Knight shift measurement \cite{matano16}. The above twofold in-plane rotational symmetry is confirmed further by field-angle dependent specific heat measurement \cite{yonezawa17}. These new experiments indicate that the SC pairing in Cu$_{x}$Bi$_{2}$Se$_{3}$ might be the first example of a pairing breaking spontaneously the spin rotation symmetry of the parent material, from the threefold rotational symmetry of the normal phase to the uniaxial twofold in-plane symmetry \cite{matano16,fu16}.

Amazingly, the new NMR experiment is claimed to be explainable by an odd-parity pairing proposed earlier \cite{fu10}. The alluded pairing, with a salient in-plane anisotropy, has also been called a nematic SC \cite{fu14}. For a simplified model with spheroidal Fermi surface, this SC pairing is known to be equivalent to the $A$ phase of $^{3}$He  and has two bulk nodes \cite{yip13,hao14}. However, by including in the model the terms responsible for hexagonal warping of the Fermi surface, this SC was argued to have a full pairing gap \cite{fu14}. If confirmed, this could be the first three-dimensional topological superconductor with a fully-gapped bulk ever discovered. However, a hallmark of the topological SC is the presence of surface Andreev bound states (SABSs) within the bulk SC gap, and its robustness against nonmagnetic impurities and interactions. In early studies in terms of simplified models based on spheroidal Fermi surface without the hexagonal warping terms, the SC is known to support novel SABSs which are flat along one direction [i.e., $(k_{x},0)$] and linearly dispersive in the perpendicular direction [i.e., $(0,k_{y})$] \cite{hao11,sasaki11,yamakage12}. In addition, the topological surface states (TSSs) are well defined at the chemical potential and well separated from the bulk states\cite{wray10}. The SC pairing would not open a gap in the TSSs and thus the TSSs coexist with the SABSs \cite{hao11,hao15}. If the hexagonal warping term is incorporated, then on one hand the two bulk nodes would be gapped out \cite{fu14}, whereas on the other hand the TSSs will remain gapless \cite{hao15}. In addition, the odd-parity topological SC pairing should still supports SABSs, independent of the model parameters \cite{hao14,yip13}. As a result, it seems natural to expect the existence of the nontrivial spectral features related to the two types of surface states.

Previous experiments, on the other hand, have made conflicting statements on the surface states in the SC phase. In several point contact spectroscopy (PCS) studies, a pronounced zero-bias peak appeared and was ascribed to the existence of SABSs \cite{sasaki11,koren11,kirzhner12}. Later, a scanning tunneling spectroscopy (STS) study reported a standard BCS-like spectrum \cite{levy12}. Moreover, a detailed PCS experiment arrived at the same conclusion that no SABSs existed \cite{peng13}. However, recent experiments clearly indicated that the Fermi surface of the Cu$_{x}$Bi$_{2}$Se$_{3}$ compound changes from spheroidal to corrugated-cylindrical surfaces as the doping $x$ increases \cite{lahoud13}. As a result and without detailed investigations, there exist works \cite{lahoud13,fu14,mizushima14} arguing that the absence of SABSs in the odd-parity (e.g., nematic) SC state of Cu$_{x}$Bi$_{2}$Se$_{3}$ is consistent with a (corrugated) cylindrical Fermi surface.

Inspired by the above experimental observations and theoretical arguments, in this work we explore whether the corrugated cylindrical Fermi surface would or would not support the SABSs in the Cu$_{x}$Bi$_{2}$Se$_{3}$ compound with the nematic SC pairing proposed for explaining the recent experiments \cite{matano16,yonezawa17}. Using the band parameters which fit qualitatively the experimental Fermi surfaces and the TSSs, we find that the SC is fully gapped, and the bulk gap is rather anisotropic. The minima of the bulk SC gap is smaller than the maximum of the bulk gap by two to three orders of magnitude. The bulk quasiparticle spectrum, while in principle is fully gapped, appears to be nodal-like from the point of view of measurements. The SABSs are shown to exist in the clean system for both spheroidal and corrugated cylindrical Fermi surfaces, and the zero-energy Majorana bound state is a characteristic of the SABSs. In addition, we verify the stability of the SABSs against short-range nonmagnetic impurities, both for uniformly distributed bulk impurities and for dilute impurities doped only to the surface layer. As to whether the SABSs could be observed experimentally may depend on the condition of the sample surfaces. For instance, the excessive magnetic Cu$^{2+}$ ($3d^9$) ions or Cu ($3d^{10}4S^{1}$) atoms on the surfaces could very much suppress the SABSs. For clean and perfect surface, the SABSs should be detectable. Recently, the possible existence of surface states in superconducting Sr$_{x}$Bi$_{2}$Se$_{3}$ is inferred from the Shubnikov-de Hass oscillation measurement \cite{liu15}. In combination with the results of the present work, it is therefore highly desirable to reexamine the existence of the SABSs in Cu$_{x}$Bi$_{2}$Se$_{3}$ as a crucial test for the relevancy of the proposed SC pairing \cite{matano16,fu14,nagai15}.

\section{model}

We consider a two-orbital tight-binging model for the low-energy degrees of freedom of the material
\begin{eqnarray}
H_{0}(\mathbf{k})&=&\epsilon(\mathbf{k})I_{4}+M(\mathbf{k})\Gamma_5+B_{0}c_{z}(\mathbf{k})\Gamma_{4}
+A_{0}[c_{y}(\mathbf{k})\Gamma_{1}   \notag \\
&&-c_{x}(\mathbf{k})\Gamma_{2}]+R_{1}d_{1}(\mathbf{k})\Gamma_{3}+R_{2}d_{2}(\mathbf{k})\Gamma_{4}.
\end{eqnarray}
The basis vector is taken as $\phi^{\dagger}_{\mathbf{k}}=[a^{\dagger}_{\mathbf{k}\uparrow}, a^{\dagger}_{\mathbf{k}\downarrow}, b^{\dagger}_{\mathbf{k}\uparrow}, b^{\dagger}_{\mathbf{k}\downarrow}]$, where the two orbitals ($a$ and $b$) are mainly from the two $p_{z}$ orbitals on the top and bottom Se layers of each Bi$_{2}$Se$_{3}$ quintuple unit \cite{zhang09,fu09,liu10}. $I_4$ is the $4\times4$ unit matrix. $\Gamma_1=\sigma_{3}\otimes s_{1}$, $\Gamma_2=\sigma_{3}\otimes s_{2}$, $\Gamma_3=\sigma_{3}\otimes s_{3}$,
$\Gamma_{4}=-\sigma_{2}\otimes s_{0}$, and $\Gamma_{5}=\sigma_{1}\otimes s_{0}$ \cite{wang10,hao11,fu09,fu10,sasaki11,liu10,zhang09}.
$s_{i}$ and $\sigma_{i}$ ($i=1,2,3$) are Pauli matrices for the spin and orbital degrees of freedom. The parity operator is defined as $P=\sigma_{1}\otimes s_{0}$ \cite{fu07}. The above model was obtained previously based on symmetry and comparison with an existing $\mathbf{k}\cdot\mathbf{p}$ model \cite{hao13,liu10}. The lattice of Bi$_2$Se$_3$ and Cu$_{x}$Bi$_2$Se$_3$, which belong to the $D_{3d}^{5}$ space group, is mapped to a hexagonal lattice in the tight-binding model. The in-plane (labeled as the $xy$ plane) and out-of-plane (labeled as the $z$ direction) lattice parameters, $a$ and $c$, are taken as $a$=4.14 \text{\AA} and $3c$=28.64 \text{\AA} \cite{acparameters}. $\epsilon(\mathbf{k})=C_{0}+2C_{1}[1-\cos(\mathbf{k}\cdot\boldsymbol{\delta}_{4})]
+\frac{4}{3}C_{2}[3-\cos(\mathbf{k}\cdot\boldsymbol{\delta}_{1})-\cos(\mathbf{k}\cdot\boldsymbol{\delta}_{2})
-\cos(\mathbf{k}\cdot\boldsymbol{\delta}_{3})]$. $M(\mathbf{k})$ is obtained from $\epsilon(\mathbf{k})$
by making the substitutions $C_{i}\rightarrow M_{i} (i=0,1,2)$. $c_{x}(\mathbf{k})=\frac{1}{\sqrt{3}}
[\sin(\mathbf{k}\cdot\boldsymbol{\delta}_{1})-\sin(\mathbf{k}\cdot\boldsymbol{\delta}_{2})]$,
$c_{y}(\mathbf{k})=\frac{1}{3}[\sin(\mathbf{k}\cdot\boldsymbol{\delta}_{1})+\sin(\mathbf{k}\cdot\boldsymbol{\delta}_{2})
-2\sin(\mathbf{k}\cdot\boldsymbol{\delta}_{3})]$, and $c_{z}(\mathbf{k})=\sin(\mathbf{k}\cdot\boldsymbol{\delta}_{4})$.
Finally, $d_{1}(\mathbf{k})=-\frac{8}{3\sqrt{3}}[\sin(\mathbf{k}\cdot\mathbf{a}_{1})+\sin(\mathbf{k}\cdot\mathbf{a}_{2})
+\sin(\mathbf{k}\cdot\mathbf{a}_{3})]$ and $d_{2}(\mathbf{k})=-8[\sin(\mathbf{k}\cdot\boldsymbol{\delta}_{1})
+\sin(\mathbf{k}\cdot\boldsymbol{\delta}_{2})+\sin(\mathbf{k}\cdot\boldsymbol{\delta}_{3})]$. Here, the four nearest-neighboring bond vectors of the hexagonal lattice are $\boldsymbol{\delta}_{1}=(\frac{\sqrt{3}}{2}a, \frac{1}{2}a, 0)$,
$\boldsymbol{\delta}_{2}=(-\frac{\sqrt{3}}{2}a$, $\frac{1}{2}a, 0)$, $\boldsymbol{\delta}_{3}=(0, -a, 0)$, and $\boldsymbol{\delta}_{4}=(0, 0, c)$.
The three in-plane second-nearest-neighboring bond vectors in $d_{1}(\mathbf{k})$ are $\mathbf{a}_{1}=\boldsymbol{\delta}_{1}-\boldsymbol{\delta}_{2}$,
$\mathbf{a}_{2}=\boldsymbol{\delta}_{2}-\boldsymbol{\delta}_{3}$, and $\mathbf{a}_{3}=\boldsymbol{\delta}_{3}-\boldsymbol{\delta}_{1}$. The last two terms in $H_{0}(\mathbf{k})$ induce hexagonal warping of the Fermi surface and the topological surface states (TSSs)\cite{fu09,liu10}.

Before doping with copper, the Fermi surface of Bi$_2$Se$_3$ is spheroidal. After intercalating copper to inter-quintuple-layer positions, the material becomes more two-dimensional. According to the experiments \cite{lahoud13}, the Fermi surface for certain Cu$_{x}$Bi$_{2}$Se$_{3}$ becomes (corrugated) cylindrical, although the details of the evolution are still unclear. On the other hand, a common feature of the normal phase of superconducting Cu$_{x}$Bi$_{2}$Se$_{3}$ is that the TSSs are well defined and coexist with the Fermi surface \cite{wray10,lahoud13}. In this work, we consider three sets of parameters shown in Table I. The Fermi surface contours (on the $k_{y}=0$ plane) and the surface spectral functions for the three sets of parameters are shown in Figure 1. The surface spectral functions are calculated in terms of the iterative Green's function method \cite{wang10,hao11,hao13}, for the upper $xy$ surface of a sample that can be regarded as consisting of an infinite number of layers. The parameters are chosen here to fit qualitatively three different shapes of the Fermi surfaces and the coexisting TSSs. The second corrugated cylindrical Fermi surface is less corrugated compared to the first corrugated cylindrical Fermi surface. Therefore, the material described by the `Cylindrical 2' is more two-dimensional than the material described by the `Cylindrical 1'. With the three typical sets of parameters, we can study the qualitative evolution of the property of a pairing as the Fermi surface turn from spheroidal to corrugated cylindrical and then becomes even more two-dimensional.

Note that, these parameter sets are chosen to reflect the evolution of the Fermi surface and the coexistence with the TSSs, which are most crucial for the low-energy physics in the superconducting phase. A completely two-dimensional model with zero hopping along the $z$ direction is unsuitable because it cannot give the TSSs observed in experiments \cite{wray10,lahoud13}. In addition, the relative magnitudes of the various parameters in Table I are in agreement with the set of parameters obtained previously by fitting the first-principle band structures for Bi$_{2}$Se$_{3}$ \cite{hao13,liu10}. By increasing the value of $R_{1}$ artificially (e.g., to 2 eV) and keeping other parameters unchanged, the topology of the Fermi surface and the coexisting TSSs can still be retained qualitatively. We will discuss the effects of increasing $R_{1}$ at the end of Section V-B.

\begin{table}[ht]
\caption{Three parameter sets for the tight-binding model, in units of electron volts (eV). `Spheroidal' and `Cylindrical' refer to the shape of the Fermi surface, which are realized with a chemical potential $\mu=0.32$ eV, for example. Two different cylindrical Fermi surface (labeled by 1 and 2) are considered.} \centering
\begin{tabular}{c c c c c c}
\hline\hline
$$ & $C_{0}$ & $C_{1}$ & $C_{2}$ & $M_{0}$ & $M_{1}$\\ [0.2ex]
\hline
Spheroidal & -0.008 & 0.06 & 1 & -0.26 & 0.3\\
\hline
Cylindrical 1 & -0.008 & 0.02 & 0.5 & -0.26 & 0.12\\
\hline
Cylindrical 2 & -0.008 & 0.02 & 0.5 & -0.26 & 0.1\\
\hline  
$$ & $M_{2}$ & $A_{0}$ & $B_{0}$ & $R_{1}$ & $R_{2}$ \\ [0.2ex]
\hline
Spheroidal & 1.2 & 0.8 & 0.35 & 0.2 & -0.3 \\
\hline
Cylindrical 1 & 0.6 & 0.6 & 0.22 & 0.2 & -0.3 \\
\hline
Cylindrical 2 & 0.6 & 0.6 & 0.19 & 0.2 & -0.3 \\
\hline
\hline
\end{tabular}
\end{table}

\begin{figure}\label{fig1} \centering
\hspace{-2.95cm} {\textbf{(a)}} \hspace{3.8cm}{\textbf{(b)}}\\
\hspace{0cm}\includegraphics[width=4.3cm,height=3.6cm]{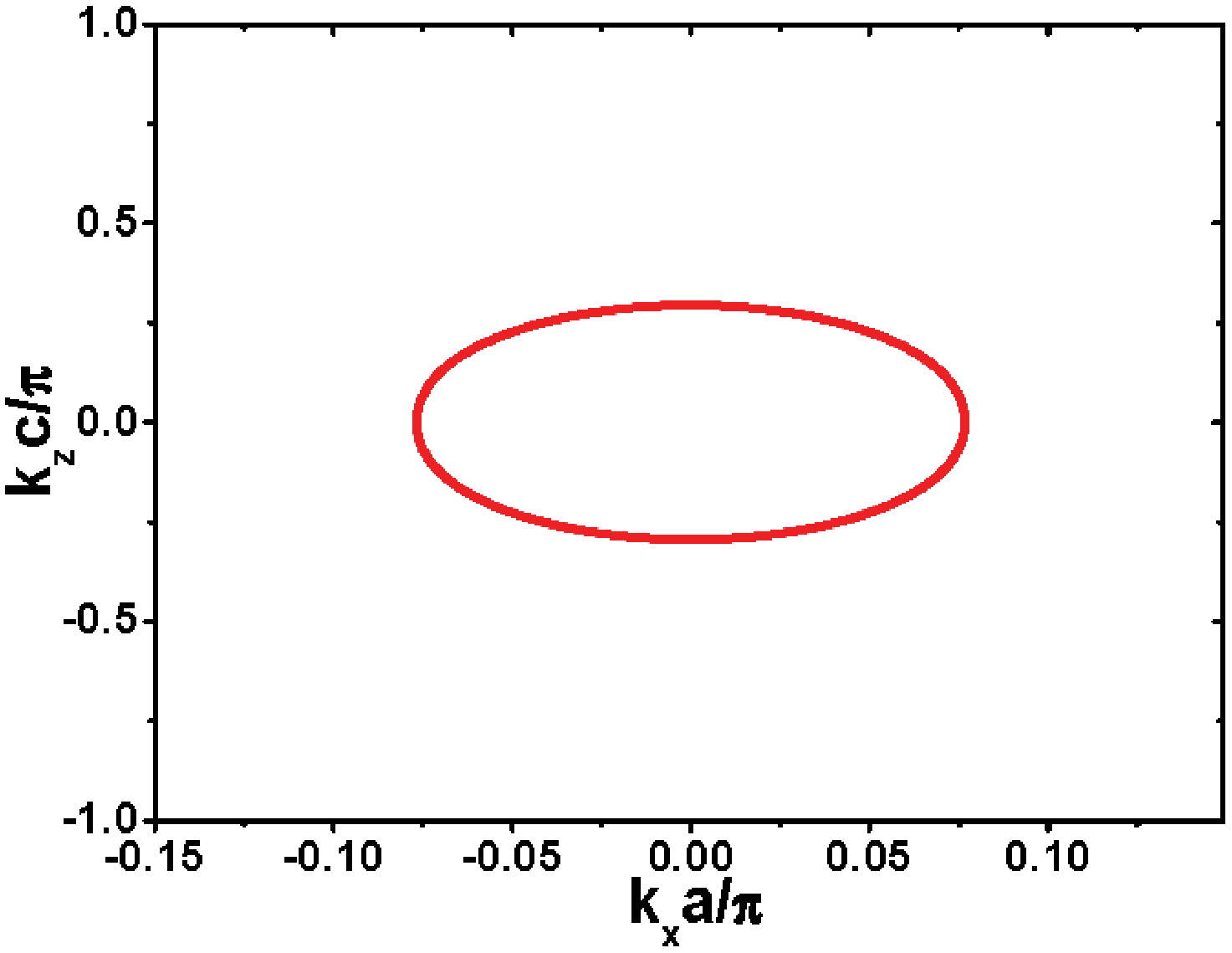}
\includegraphics[width=4.2cm,height=3.5cm]{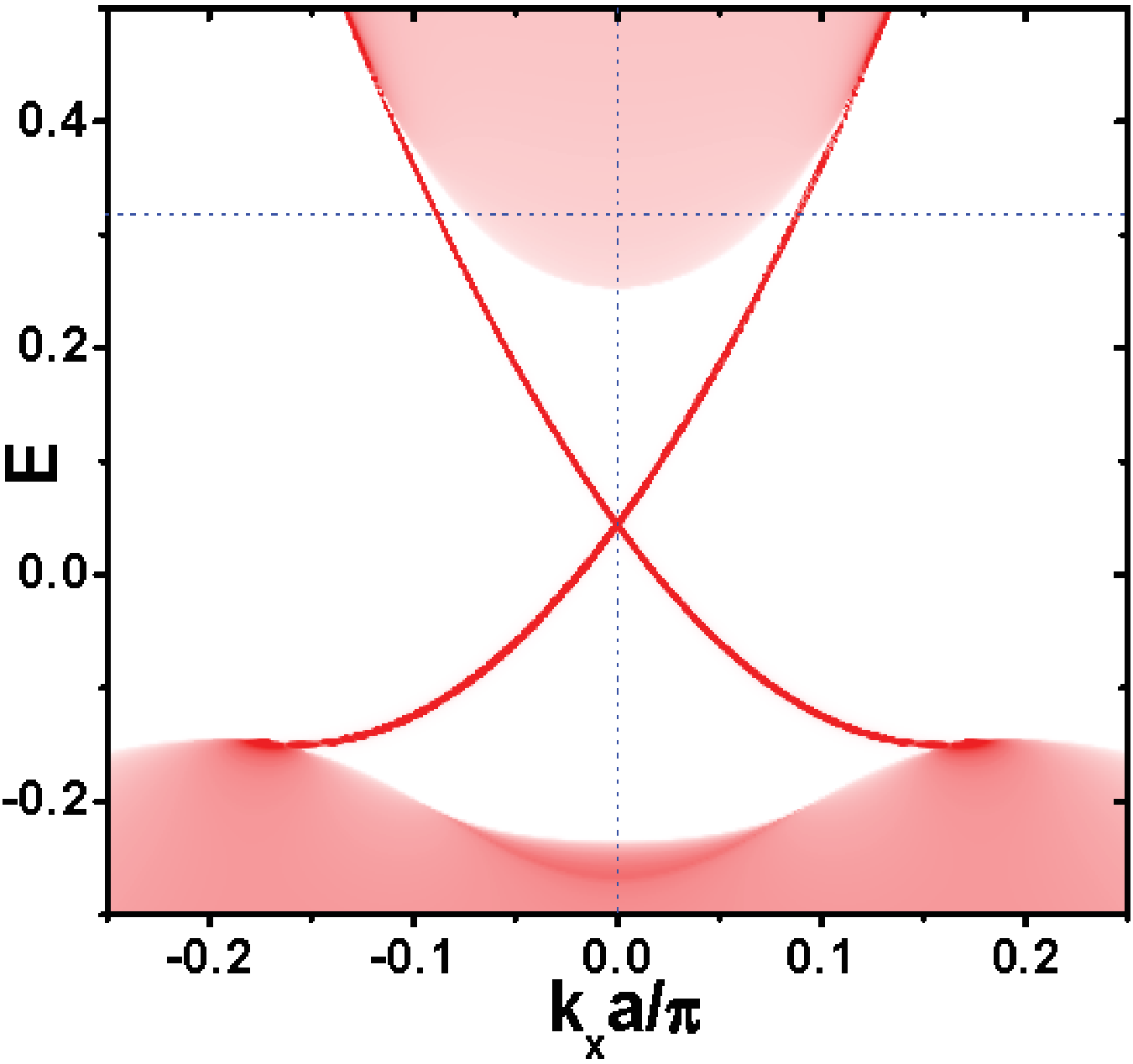}\\
\vspace{-0.10cm}
\hspace{-2.95cm} {\textbf{(c)}} \hspace{3.8cm}{\textbf{(d)}}\\
\hspace{0cm}\includegraphics[width=4.3cm,height=3.6cm]{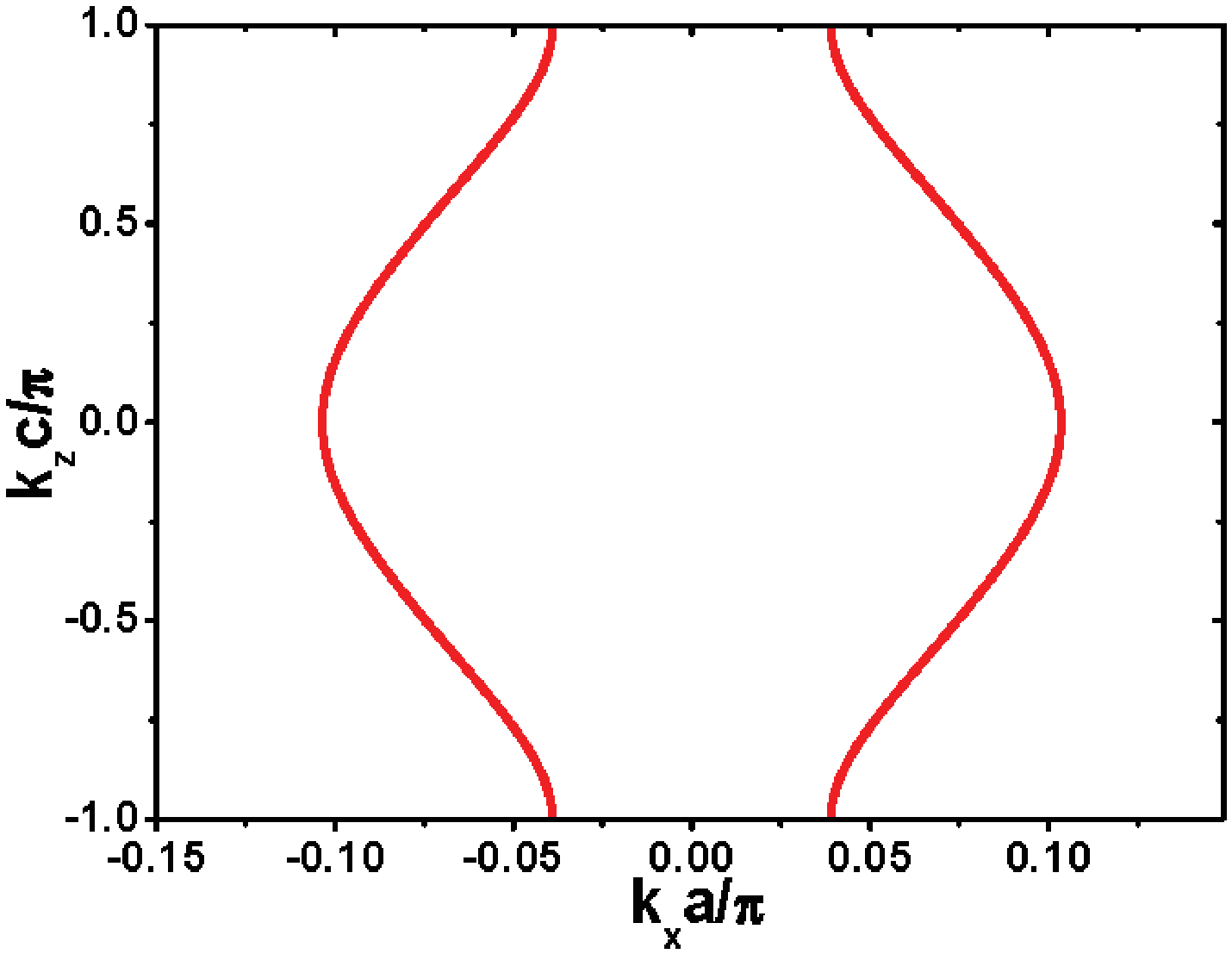}
\includegraphics[width=4.2cm,height=3.5cm]{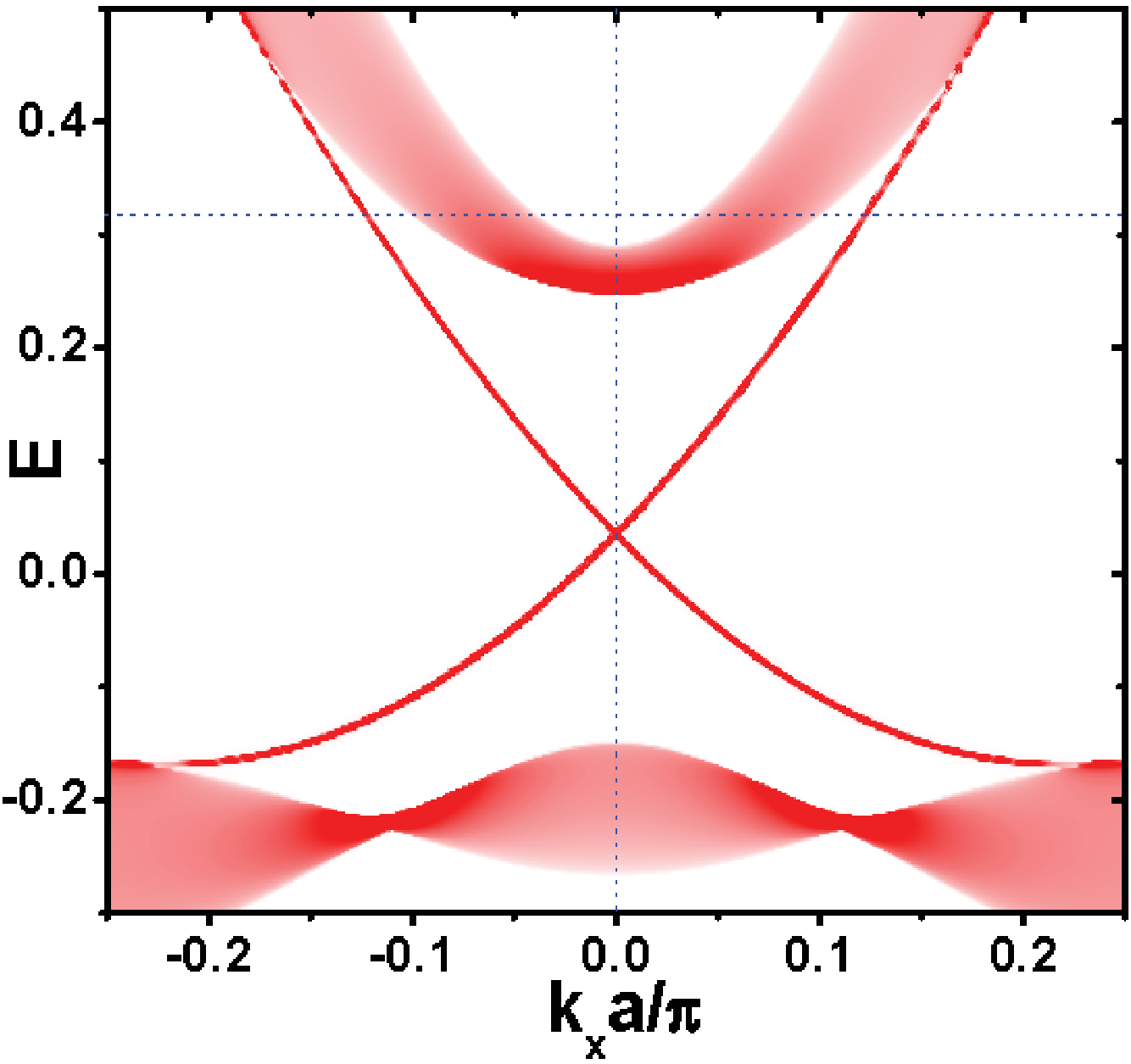}   \vspace{-0.10cm} \\
\hspace{-2.95cm} {\textbf{(e)}} \hspace{3.8cm}{\textbf{(f)}}\\
\hspace{0cm}\includegraphics[width=4.3cm,height=3.6cm]{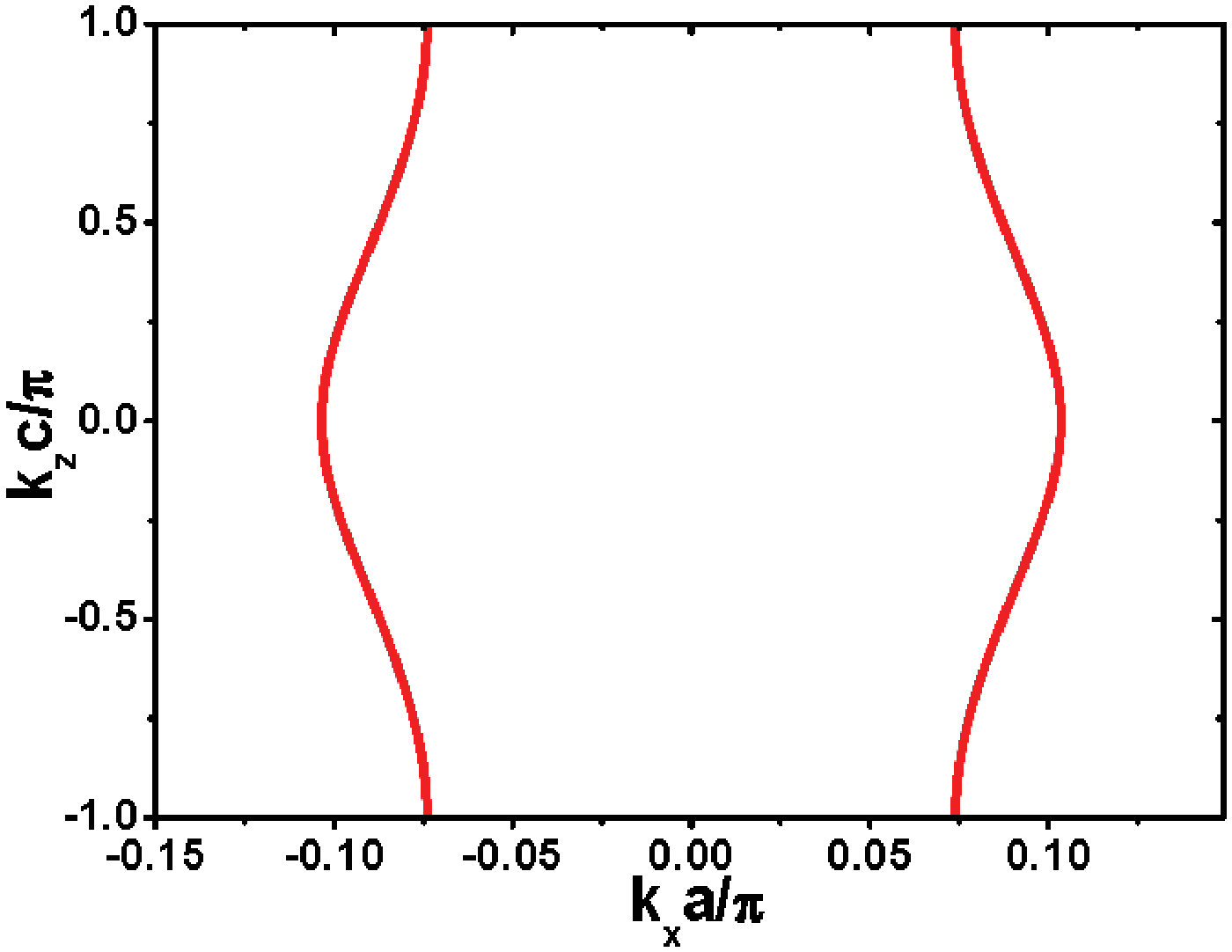}
\includegraphics[width=4.2cm,height=3.5cm]{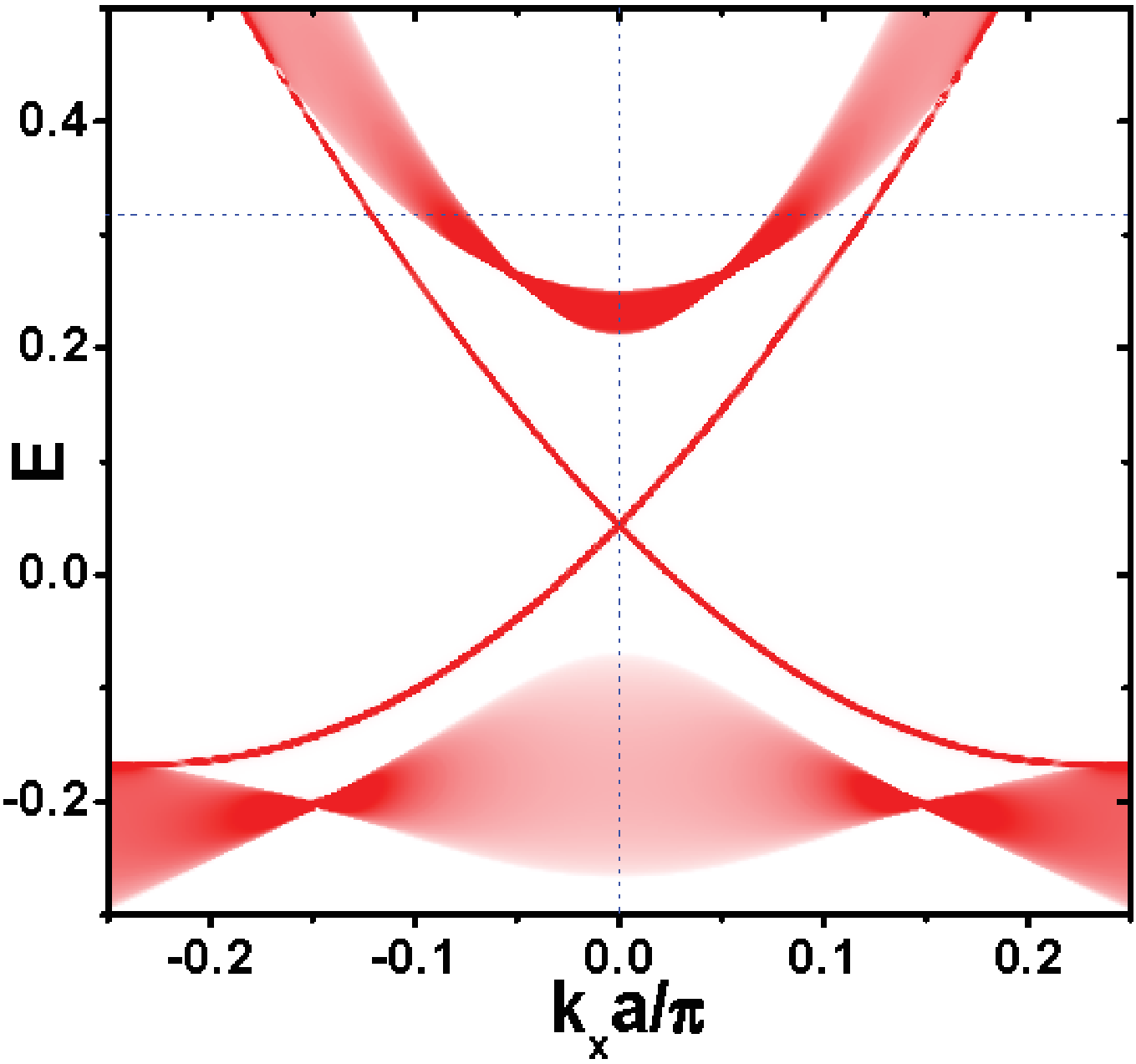}
\caption{(Color online) (a,c,e) Cross-sections of the Fermi surface with the $k_{y}=0$ plane. (b,d,f) The surface spectral functions on the upper $xy$ surface of a thick film. (a) and (b) are for the first set of parameters (`Spheroidal') in Table I. (c) and (d) are for the second set of parameters (`Cylindrical 1') in Table I. (e) and (f) are for the third set of parameters (`Cylindrical 2') in Table I. The energy $E$ is in unit of electron volts. $\mu=0.32$ eV. The horizontal dotted lines in (b,d,f) mark the Fermi level.}
\end{figure}

\section{Pseudospin basis}

The full model contains the complete information of the phase, but it is hard to work with analytically. On the other hand, it is the states close to the Fermi level that are most important to the superconducting phase. By introducing the pseudospin basis, the full model containing both of the two bands of the model in the normal phase can be projected to a simplified model containing only the band contributing to the Fermi surface  \cite{yip13,yip14,yip16,zocher13,hashimoto13,nagai14,takami14}. By making this reduction, the low-energy properties of the superconducting phase, in particular the gap structure of the bulk quasiparticle spectrum and the SABSs, can be understood more easily. Here, we follow the approach of Yip, which was originally applied to a simplified version of the model, to construct the pseudospin basis for our tight-binding model \cite{yip13,yip16}. This method makes use of the time-reversal symmetry ($T$) and inversion symmetry ($P$) of the model, which lead to the Kramers degeneracy of each state. The two pseudospin bases for each Kramers doublet are thus required to be related by the combined action of $PT$ operation. Throughout this work, we assume the chemical potential to lie within the conduction band. The eigenbasis can be constructed by first diagonalizing the model in the spin subspace and then in the orbital subspace. One basis can be taken as
\begin{equation}
|\mathbf{k},\alpha'\rangle=\frac{1}{\tilde{D}_{\mathbf{k}}N_{\mathbf{k}}}\begin{pmatrix} \tilde{E}_{\mathbf{k}} \\ \tilde{M}_{-}(\mathbf{k}) \end{pmatrix} \begin{pmatrix} A_{0}c_{+}(\mathbf{k}) \\ D_{-}(\mathbf{k}) \end{pmatrix},
\end{equation}
where the first and second vectors are separately spinors in the subspaces of the original orbital and spin degree of freedom. For notational simplicity, here and later in this work we will use the following abbreviations
$c_{\pm}(\mathbf{k})=c_{y}(\mathbf{k})\pm i c_{x}(\mathbf{k})$, $\tilde{M}_{\pm}(\mathbf{k})=M(\mathbf{k})\pm i[B_{0}c_{z}(\mathbf{k})+R_{2}d_{2}(\mathbf{k})]$, $D_{\mathbf{k}}=\sqrt{A_{0}^{2}[c_{x}^{2}(\mathbf{k})+c_{y}^{2}(\mathbf{k})]+R_{1}^{2}d_{1}^{2}(\mathbf{k})}$, $E_{\mathbf{k}}=\sqrt{|\tilde{M}_{\pm}(\mathbf{k})|^{2}+D_{\mathbf{k}}^{2}}$, $\tilde{E}_{\mathbf{k}}=E_{\mathbf{k}}+D_{\mathbf{k}}$, $N_{\mathbf{k}}=\sqrt{2E_{\mathbf{k}}\tilde{E}_{\mathbf{k}}}$, $D_{\pm}(\mathbf{k})=D_{\mathbf{k}}\pm R_{1}d_{1}(\mathbf{k})$, $\tilde{D}_{\mathbf{k}}=\sqrt{2D_{\mathbf{k}}D_{-}(\mathbf{k})}$. The other pseudospin basis is related to the one listed above by symmetry
\begin{equation}
|\mathbf{k},\beta'\rangle=PT|\mathbf{k},\alpha'\rangle=\frac{1}{\tilde{D}_{\mathbf{k}}N_{\mathbf{k}}}\begin{pmatrix} \tilde{M}_{+}(\mathbf{k})  \\ \tilde{E}_{\mathbf{k}} \end{pmatrix} \begin{pmatrix} -D_{-}(\mathbf{k})  \\ A_{0}c_{-}(\mathbf{k})   \end{pmatrix},
\end{equation}
In order for the pseudospin basis to have the correct $P$ and $T$ symmetries in the whole BZ, we fix the wave vectors in $|\mathbf{k},\alpha'\rangle$ and $|\mathbf{k},\beta'\rangle$ to lie on the northern hemisphere ($k_{z}>0$). States on the southern hemisphere are obtained by symmetry, namely $|-\mathbf{k},\alpha'\rangle=P|\mathbf{k},\alpha'\rangle$ and $|-\mathbf{k},\beta'\rangle=P|\mathbf{k},\beta'\rangle=T|\mathbf{k},\alpha'\rangle$.

The naive choice of the pseudospin basis defined above are not guaranteed to have the correct rotational property of the original model. As a result, they may not be the suitable basis set for studying the symmetry of a specific pairing channel. As has been shown in Ref.\cite{yip16}, a good set of the pseudospin basis can be constructed as a linear combination of $|\mathbf{k},\alpha'\rangle$ and $|\mathbf{k},\beta'\rangle$ that make the magnetic moment operator expressed under this basis to have the proper transformation property under rotation \cite{yip13}. For the model defined by Eq.(1), the magnetic moment is a linear combination of $\mathbf{s}$ and $\sigma_{1}\mathbf{s}$. Here, following the method in Ref.\cite{yip16}, we choose to focus on the component $m_{1\alpha}=g_{1\alpha}\frac{\sigma_{0}+\sigma_{1}}{2}s_{\alpha}$ of the magnetic moment \cite{yip16}. $\alpha=x,y,z$, and $g_{1x}=g_{1y}=g_{1p}$ are assumed. In the space of $\{|\mathbf{k},\alpha'\rangle, |\mathbf{k},\beta'\rangle\}$, $m_{1z}$ has the following matrix form
\begin{equation}
\frac{m_{1z}(\mathbf{k})}{g_{1z}|W_{\mathbf{k}}|^{2}}=\begin{pmatrix} \cos\theta_{\mathbf{k}} & ie^{i(\varphi_{\mathbf{k}}+2\phi_{\mathbf{k}})}\sin\theta_{\mathbf{k}} \\   -ie^{-i(\varphi_{\mathbf{k}}+2\phi_{\mathbf{k}})}\sin\theta_{\mathbf{k}}  &  -\cos\theta_{\mathbf{k}}  \end{pmatrix},
\end{equation}
where the three phase factors, $\varphi_{\mathbf{k}}$, $\phi_{\mathbf{k}}$, and $\theta_{\mathbf{k}}$, are defined by
\begin{equation}
c_{+}(\mathbf{k})=i\sqrt{c_{x}^{2}(\mathbf{k})+c_{y}^{2}(\mathbf{k})}e^{-i\varphi_{\mathbf{k}}}=ic(\mathbf{k})e^{-i\varphi_{\mathbf{k}}},
\end{equation}
\begin{equation}
W_{\mathbf{k}}=\frac{\tilde{E}_{\mathbf{k}}+\tilde{M}_{+}(\mathbf{k})}{\sqrt{2}N_{\mathbf{k}}}=|W_{\mathbf{k}}|e^{i\phi_{\mathbf{k}}},
\end{equation}
\begin{equation}
R_{1}d_{1}(\mathbf{k})+iA_{0}c(\mathbf{k})=D_{\mathbf{k}}e^{i\theta_{\mathbf{k}}}.
\end{equation}
$m_{1z}(\mathbf{k})$ in the basis of  $\{$$|\mathbf{k},\alpha'\rangle$, $|\mathbf{k},\beta'\rangle$$\}$ clearly does not have the desired form of the $z$-component of an axial vector. The new basis $|\mathbf{k},\alpha\rangle$ and $|\mathbf{k},\beta\rangle$ are constructed such that $m_{1z}(\mathbf{k})$ is proportional to the $z$-component of the Pauli matrix, namely they are the eigenbasis of $m_{1z}$ \cite{yip13,yip16}. We take
\begin{equation}
|\mathbf{k},\alpha\rangle=h(\mathbf{k})[(1+\cos\theta_{\mathbf{k}})|\mathbf{k},\alpha'\rangle -ie^{-i(\varphi_{\mathbf{k}}+2\phi_{\mathbf{k}})}\sin\theta_{\mathbf{k}}|\mathbf{k},\beta'\rangle].
\end{equation}
From $|\mathbf{k},\beta\rangle=PT|\mathbf{k},\alpha\rangle$, we get the other basis
\begin{equation}
|\mathbf{k},\beta\rangle=h^{\ast}(\mathbf{k})[(1+\cos\theta_{\mathbf{k}})|\mathbf{k},\beta'\rangle -ie^{i(\varphi_{\mathbf{k}}+2\phi_{\mathbf{k}})}\sin\theta_{\mathbf{k}}|\mathbf{k},\alpha'\rangle].
\end{equation}
Normalization of the eigenbasis requires
\begin{equation}
|h(\mathbf{k})|^{2}=\frac{1}{2(1+\cos\theta_{\mathbf{k}})}=\frac{1}{4\cos^{2}\frac{\theta_{\mathbf{k}}}{2}}.
\end{equation}
In this basis, we have $m_{1z}(\mathbf{k})=g_{1z}|W_{\mathbf{k}}|^{2}\rho_{z}$, where $\rho_{z}$ is the conventional $z$-component of the Pauli matrices. The $x$-component of $\mathbf{m}_{1}$ in the new basis has a purely off-diagonal form with the two off-diagonal elements
\begin{equation}
\frac{[m_{1x}(\mathbf{k})]_{\alpha\beta}}{g_{1p}|W_{\mathbf{k}}|^{2}}=\frac{[m_{1x}(\mathbf{k})]^{\ast}_{\beta\alpha}}{g_{1p}|W_{\mathbf{k}}|^{2}}=
[2ih^{\ast}(\mathbf{k})\cos\frac{\theta_{\mathbf{k}}}{2}e^{i(\varphi_{\mathbf{k}}+\phi_{\mathbf{k}})}]^{2}.
\end{equation}
Taking
\begin{equation}
h(\mathbf{k})=\frac{i}{2\cos\frac{\theta_{\mathbf{k}}}{2}}e^{i(\varphi_{\mathbf{k}}+\phi_{\mathbf{k}})},
\end{equation}
we have
\begin{equation}
m_{1x}(\mathbf{k})=g_{1p}|W_{\mathbf{k}}|^{2}\rho_{x},
\end{equation}
\begin{equation}
m_{1y}(\mathbf{k})=g_{1p}|W_{\mathbf{k}}|^{2}\rho_{y},
\end{equation}
where $\rho_{x}$ and $\rho_{y}$ are the conventional $x$-component and $y$-component of the Pauli matrices. Therefore, we have shown that the new basis $\{$$|\mathbf{k},\alpha\rangle$, $|\mathbf{k},\beta\rangle$$\}$ defined by Eqs.(8), (9), and (12) can ensure the correct transformation property of the magnetic moment operator, and are thus proper choices in discussing symmetry properties of the system. This basis, employed in the present work, is  shown \cite{yip16} to coincide with the so-called manifestly covariant Bloch basis introduced by Fu \cite{fu15,kozii15,venderbos16a,venderbos16b}.

\section{pairing and gap structure of the bulk quasiparticle spectrum}

In the Nambu basis, $\psi^{\dagger}_{\mathbf{k}}=[\phi^{\dagger}_{\mathbf{k}},(\phi_{-\mathbf{k}})^{\text{T}}]$, and denoting the pairing term generically as $\underline{\Delta}(\mathbf{k})$, the model for a bulk superconducting topological insulator is written as
\begin{eqnarray}
\hat{H}&=&\frac{1}{2}\sum\limits_{\mathbf{k}}\psi^{\dagger}_{\mathbf{k}}\begin{pmatrix} H_{0}(\mathbf{k})-\mu I_{4} & \underline{\Delta}(\mathbf{k}) \\
-\underline{\Delta}^{\ast}(-\mathbf{k}) & \mu I_{4}-H^{\ast}_{0}(-\mathbf{k})
\end{pmatrix}\psi_{\mathbf{k}}   \notag \\
&=&\frac{1}{2}\sum\limits_{\mathbf{k}}\psi^{\dagger}_{\mathbf{k}}H(\mathbf{k})\psi_{\mathbf{k}},
\end{eqnarray}
where $\mu$ is the chemical potential. The $\frac{1}{2}$ factor accounts for the particle-hole redundancy introduced by the Nambu representation.

The two-fold in-plane rotation symmetry in the Knight shift and field-angle dependent specific heat experiments indicate that the pairing must belong to a multi-dimensional representation of the symmetry group. Because the pairing order parameter for a one-dimensional representation should necessarily respect the three-fold rotational symmetry of the $D_{3d}^{5}$ space group. Presently, most of attention has been paid to the two-dimensional $E_{u}$ representation of the $D_{3d}^{5}$ space group. One set of the two bases for the $E_{u}$ representation is $\underline{\Delta}_{4a}(\mathbf{k})=i\Delta_{a}\sigma_{2}\otimes s_{0}$ and $\underline{\Delta}_{4b}(\mathbf{k})=\Delta_{b}\sigma_{2}\otimes s_{3}$, with $\Delta_{a}$ and $\Delta_{b}$ the pairing amplitudes. For a simplified model without the hexagonal warping terms (i.e., $R_{1}=R_{2}=0$), both of the two components are known to lead to bulk spectrum with point nodes \cite{fu10,hao11}. However, it was shown by Fu that the bulk nodes for $\underline{\Delta}_{4a}(\mathbf{k})$ are gapped out by including the hexagonal warping term proportional to $R_{1}$ of Eq.(1) \cite{fu14}. It seems that the fully gapped $\underline{\Delta}_{4a}(\mathbf{k})$ provides a most natural explanation to the Knight shift and field-angle dependent specific heat experiments.

Here, we study more carefully the excitation gap of the bulk quasiparticle spectrum. Since only the conduction band contribute to the Fermi surface, the gap structure of the quasiparticle spectrum is understood more easily from the low-energy effective model obtained by projecting the full model defined by Eq.(15) to the conduction band \cite{yip13}. The dispersion of the conduction band is $\epsilon(\mathbf{k})+E_{\mathbf{k}}$. The projection is thus achieved by replacing $H_{0}(\mathbf{k})-\mu I_{4}$ with $(\epsilon(\mathbf{k})+E_{\mathbf{k}}-\mu)I_{2}$ and transforming the pairing term expressed in the spin-orbital basis to the pseudospin basis derived in the last section. For an arbitrary pairing denoted as $\underline{\Delta}(\mathbf{k})$ in the original spin-orbital basis, its expression in the new pseudospin basis of the conduction band is
\begin{equation}
\tilde{\underline{\Delta}}(\mathbf{k})=U^{\dagger}_{\mathbf{k}}\underline{\Delta}U^{\ast}_{-\mathbf{k}},
\end{equation}
where the transformation matrix is $U_{\mathbf{k}}=[|\mathbf{k},\alpha\rangle,|\mathbf{k},\beta\rangle]$.

For $\underline{\Delta}_{4a}(\mathbf{k})=i\Delta_{a}\sigma_{2}\otimes s_{0}$, we have \cite{note1}
\begin{eqnarray}
\tilde{\underline{\Delta}}_{4a}(\mathbf{k})&=&\frac{\Delta_{0}}{E_{\mathbf{k}}}[-R_{1}d_{1}(\mathbf{k})\rho_{1} -(B_{0}c_{z}(\mathbf{k})+R_{2}d_{2}(\mathbf{k}))\rho_{2}     \notag \\
&&+A_{0}c_{y}(\mathbf{k})\rho_{3}]i\rho_{2}.
\end{eqnarray}
On the Fermi surface, $\epsilon(\mathbf{k})+E_{\mathbf{k}}-\mu=0$, the quasiparticle spectrum is determined only by the pairing term
\begin{equation}
E(\mathbf{k})=\pm|\text{det}[\tilde{\underline{\Delta}}_{4a}(\mathbf{k})]|
=\pm|\Delta_{a}|\sqrt{1-\frac{M^{2}(\mathbf{k})+A_{0}^{2}c_{x}^{2}(\mathbf{k})}{E_{\mathbf{k}}^{2}}}.
\end{equation}
Up to slight hexagonal warping induced by terms proportional to $R_{1}$ and $R_{2}$, both $M(\mathbf{k})$ and $E_{\mathbf{k}}$ are approximately symmetrical in the $k_{x}k_{y}$ plane. As a result of the $c_{x}(\mathbf{k})$ term in Eq.(18), the size of superconducting gap is smaller along the $k_{y}=0$ contour of the Fermi surface than that along the $k_{x}=0$ contour of the Fermi surface. Therefore, the bulk energy spectrum for $\underline{\Delta}_{4a}(\mathbf{k})$ has a strong anisotropy between the $k_{x}$ direction and the $k_{y}$ direction.

Let us focus on the contour of the Fermi surface on the $k_{y}=0$ plane, where the minimum of the superconducting gap is attained. Eq.(18) is written as
\begin{equation}
E(\mathbf{k})=\pm\frac{|\Delta_{a}|}{\mu-\epsilon_{\mathbf{k}}}\sqrt{R_{1}^{2}d_{1}^{2}(\mathbf{k})+B_{0}^{2}c^{2}_{z}(\mathbf{k})},
\end{equation}
where $\epsilon(\mathbf{k})+E_{\mathbf{k}}-\mu=0$ has been used. For clarity, we have $d_{1}(\mathbf{k})\simeq (k_{x}a)^{3}$ for $k_{y}=0$ and $k_{x}$ small, and $c_{z}(\mathbf{k})=\sin(k_{z}c)$. Notice that for both spheroidal and corrugated cylindrical Fermi surfaces, including those shown in Fig.1, $d_{1}(\mathbf{k})$ and $c_{z}(\mathbf{k})$ do not attain zero simultaneously. While for $R_{1}=0$ there are point nodes on the $k_{y}=0$ Fermi surface contour determined by $c_{z}(\mathbf{k})=0$, a finite $R_{1}$ removes all these nodes \cite{fu14}. One exception is a spheroidal Fermi surface with a point $\mathbf{k}=(0,0,\pi)$ on it, which marks the transition between a spheroidal Fermi surface and a corrugated cylindrical Fermi surface. For practical purpose, however, we will ignore this special case and so the bulk spectrum of $\underline{\Delta}_{4a}(\mathbf{k})$ is always fully gapped for $R_{1}\ne0$.

On the other hand, $k_{x}a$ on the Fermi surface is small for actual materials. The size of the gap for $c_{z}(\mathbf{k})=0$ and $k_{y}=0$, which grows like $(k_{x}a)^{3}$ for small $k_{x}$, is actually much smaller than $\Delta_{a}$. In all cases studied, $\mu-\epsilon(\mathbf{k})=E_{\mathbf{k}}$ has only a small variation on the Fermi surface. Therefore, the numerator of Eq.(19) determines the qualitative behavior of the superconducting gap. For simplicity and without losing generality, we focus on the $k_{x}\ge0$ and $k_{z}\ge0$ portion of the $k_{y}=0$ Fermi surface contours shown in Figs.1(a, c, and e). Each point on the chosen portion of the Fermi surface contour can then by labeled by a unique $k_{x}$. For the spheroidal Fermi surface shown in Fig.1(a), as we go along the Fermi surface contour from $(0,0,k_{z1})$ ($k_{z1}c\simeq0.29\pi$) to $(k_{x1},0,0)$ ($k_{x1}a\simeq0.077\pi$), $c_{z}(\mathbf{k})$ decreases monotonously and we have $|B_{0}c_{z}(0,0,k_{z1})|\gg|R_{1}d_{1}(k_{x1},0,0)|$. The size of the gap is thus expected to decrease monotonously as we go along the contour from $(0,0,k_{z1})$ to $(k_{x1},0,0)$. For the two cases with corrugated cylindrical Fermi surfaces, the Fermi surface contour is bounded by two points $(k_{xi},0,\pi)$ and $(k_{xf},0,0)$. $k_{xi}a\simeq0.039\pi$ ($k_{xi}a\simeq0.074\pi$) and $k_{xf}a\simeq0.103\pi$ ($k_{xf}a\simeq0.103\pi$) for Fig.1(c) [Fig.1(e)]. As we increase $k_{x}$ from $k_{xi}$ to $k_{xf}$, $k_{z}$ changes from $\pi$ to 0. As a result, we have a nonmonotonous variation of $c_{z}(\mathbf{k})$, which first increases towards $1$ as $k_{z}$ approaches $\pi/2$ and then decreases to $0$ afterwards. Because we have $|R_{1}d_{1}(k_{x},0,k_{z})/B_{0}|\ll1$ along the Fermi surface contour, we expect to get a nonmonotonous variation of the superconducting gap for corrugated cylindrical Fermi surfaces, which first increases and then decreases, with two minima at $(k_{xi},0,\pi)$ and $(k_{xf},0,0)$. In comparison to the case for Fig.1(a), the number of gap minima is doubled when the Fermi surface evolves from spheroidal to corrugated cylindrical.

To have a more quantitative understanding on the evolution of the superconducting gap explained above, we plot in Fig.2 simultaneously three functions $f_{1}=4\sqrt{R_{1}^{2}d_{1}^{2}(\mathbf{k})+B_{0}^{2}c^{2}_{z}(\mathbf{k})}$, $f_{2}=4(\mu-\epsilon_{\mathbf{k}})=4E_{\mathbf{k}}$, and $f_{3}=f_{1}/f_{2}=|E(\mathbf{k})/\Delta_{a}|$, on the $k_{z}\ge0$ and $k_{x}\ge0$ portion of the $k_{y}=0$ Fermi surface contours. The states are labeled uniquely in terms of the value of $k_{x}$. The value of $f_{3}$ gives the magnitude of the bulk gap, normalized by the pairing amplitude. For all three sets of parameters considered, the minimal values of the bulk gap are much smaller than the corresponding maximum values. For the experimental transition temperature of about 3.8 Kelvin, the pairing amplitude (i.e., $\Delta_{a}$) is of the order 1 meV. The minimal value of the bulk gap is two to three orders of magnitude smaller than the pairing amplitude. It is also interesting to notice that, when the smaller Fermi momentum of the corrugated cylindrical Fermi surface along $k_{x}$ [e.g., $k_{xi}a\simeq0.039\pi$ for Fig.2(b)] is smaller than the Fermi momentum of the spheroidal Fermi surface along $k_{x}$ [e.g., $k_{x1}a\simeq0.077\pi$ for Fig.2(a)], the minimum superconducting gap for the corrugated cylindrical Fermi surface can be smaller than the minimum superconducting gap for the spheroidal Fermi surface. From a practical point of view, and for both spheroidal and corrugated cylindrical Fermi surfaces, the minimum of the superconducting gap acts effectively as point node of the bulk spectrum. The above picture holds as long as $R_{1}$ is not extremely (e.g., two to three orders of magnitude) larger than the value used. For the parameter set of `Cylindrical 2' in table I, a two orders of magnitude larger $R_{1}$ (i.e., 20 eV) is needed to increase the minimum of the bulk gap along $k_{x}$ to the same order of magnitude to that along $k_{y}$. A further tenfold enhancement in $R_{1}$ is required to achieve the same increase in the bulk gap for the parameter set of `Cylindrical 1' in table I. These large values of $R_{1}$ are not only inconsistent with the magnitudes of other parameters but also will distort strongly the Fermi surface and thus deviate qualitatively from experiments. Therefore, the bulk spectrum for $\underline{\Delta}_{4a}(\mathbf{k})$ should be nodal-like for realistic parameters.

\begin{figure}[!htb]\label{fig2}
\centering
\hspace{-5cm} {\textbf{(a)}}\\
\includegraphics[width=6.5cm,height=4.2cm]{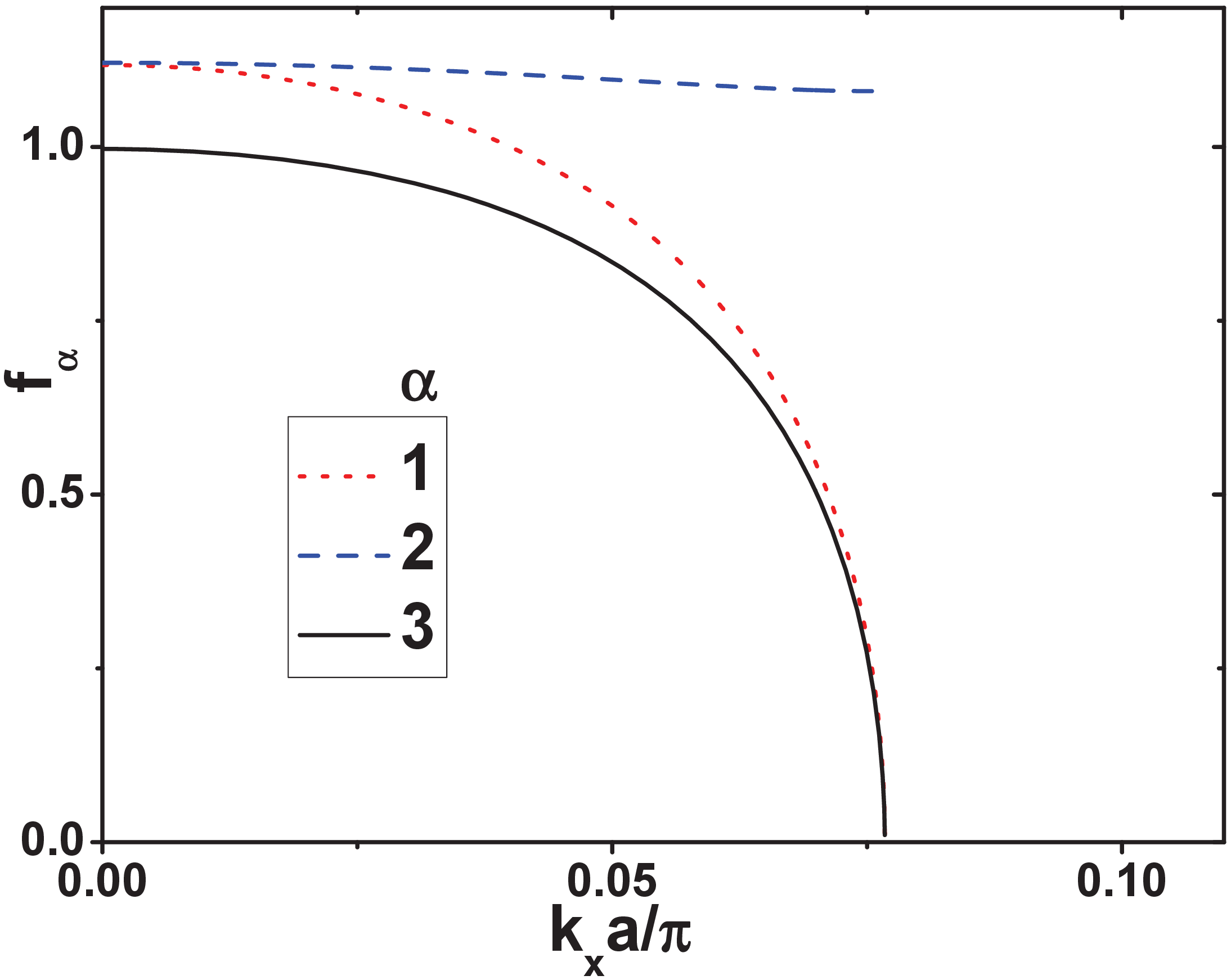} \\ \vspace{-0.05cm}
\hspace{-5cm} {\textbf{(b)}}\\
\includegraphics[width=6.5cm,height=4.2cm]{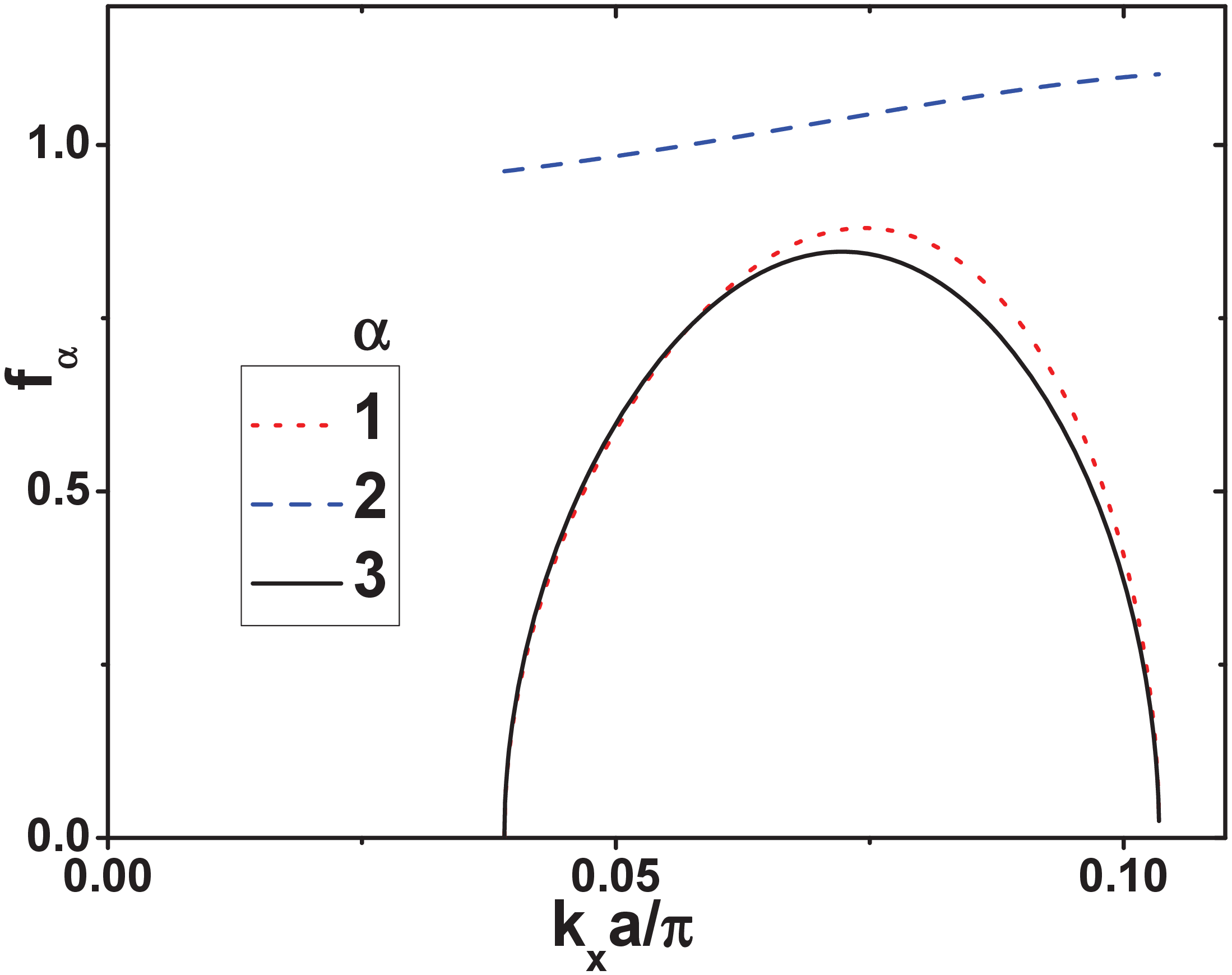}  \\ \vspace{-0.05cm}
\hspace{-5cm} {\textbf{(c)}}\\
\includegraphics[width=6.5cm,height=4.2cm]{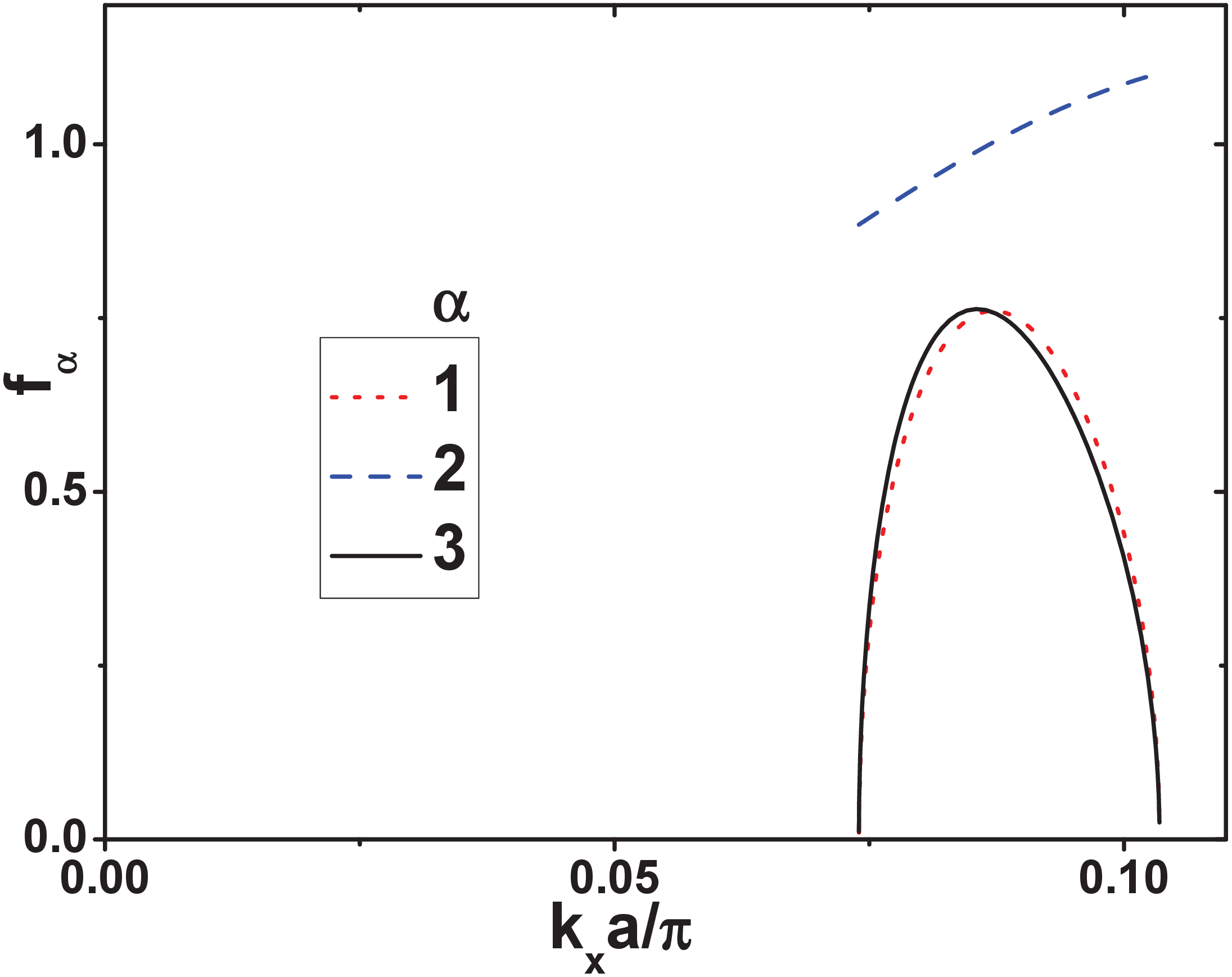}
\caption{(Color online) The evolution of the normalized superconducting gap $f_{3}$, and two functions ($f_{1}$ and $f_{2}$) determining it, as a function of $k_{x}$ along the $k_{x}\ge0$ and $k_{z}\ge0$ portion of the Fermi surface contour on the $k_{y}=0$ plane. The three functions are defined in the accompanying main text. (a), (b), and (c) are separately for the first (`Spheroidal'), second (`Cylindrical 1'), and third (`Cylindrical 2') sets of parameters in Table I. In (a), the minimal gap is $f_{3}\simeq0.0103$ for $k_{x}a\simeq0.077\pi$. In (b), the minimal gaps include $f_{3}\simeq0.0015$ for $k_{x}a\simeq0.039\pi$ and $f_{3}\simeq0.0245$ for $k_{x}a\simeq0.103\pi$. In (c), the minimal gaps include $f_{3}\simeq0.0112$ for $k_{x}a\simeq0.074\pi$ and $f_{3}\simeq0.0245$ for $k_{x}a\simeq0.103\pi$.}
\end{figure}

For $\underline{\Delta}_{4b}(\mathbf{k})=\Delta_{b}\sigma_{2}\otimes s_{3}$, the effective pairing is
\begin{eqnarray}
\tilde{\underline{\Delta}}_{4b}(\mathbf{k})&=&\frac{\Delta_{b}}{E_{\mathbf{k}}}[(B_{0}c_{z}(\mathbf{k})+R_{2}d_{2}(\mathbf{k}))\rho_{1}
-R_{1}d_{1}(\mathbf{k})\rho_{2}     \notag \\
&&-A_{0}c_{x}(\mathbf{k})\rho_{3}]i\rho_{2}.
\end{eqnarray}
The minimum of the superconducting gap lies along the $k_{x}=0$ plane, where $d_{1}(\mathbf{k})=c_{x}(\mathbf{k})=0$. Along the intersection contour of the Fermi surface with the $k_{x}=0$ plane, $E_{\mathbf{k}}=\mu-\epsilon({\mathbf{k}})$ is a smooth function of the wave vector. The variation of the gap is thus determined by $B_{0}c_{z}(\mathbf{k})+R_{2}d_{2}(\mathbf{k})$. For $R_{2}=0$, we reproduce the known result that $\underline{\Delta}_{4b}(\mathbf{k})$ has bulk point nodes determined by $c_{z}(\mathbf{k})=k_{x}=0$. The number of the point nodes is two (four) for spheroidal (corrugated cylindrical) Fermi surface. For $R_{2}\ne0$, the above point nodes are gapped out, with the magnitude of the gap proportional to $|\Delta_{b}R_{2}d_{2}(0,k_{Fy},0)|/E_{\mathbf{k}}$ [and also $|\Delta_{b}R_{2}d_{2}(0,k'_{Fy},\pi)|/E_{\mathbf{k}}$ for corrugated cylindrical Fermi surface], where $k_{Fy}$ ($k'_{Fy}$) is the $k_{y}$ component of the Fermi momentum. The original point nodes do not simply vanish. Instead, they are tilted away from the $(0,k_{y},0)$ axis [and also the $(0,k_{y},\pi)$ axis for the case with corrugated cylindrical Fermi surface] into the $k_{y}k_{z}$ plane. If we have $|R_{2}/B_{0}|\gg1$, a fully-gapped bulk spectrum can be obtained. However, for realistic parameters, the point nodes are still present. For the parameters considered in Table I, the point nodes of $\underline{\Delta}_{4b}(\mathbf{k})$ are in fact still very close to the point nodes for $R_{2}=0$.

\section{surface Andreev bound states}

Since $\underline{\Delta}_{4b}(\mathbf{k})$ is nodal for practical model parameters, we will focus on the fully-gapped $\underline{\Delta}_{4a}$ in what follows. For spheroidal Fermi surface, $\underline{\Delta}_{4a}$ was known to support a peculiar surface Andreev bound states (SABSs) on the $xy$ surface of a sample, which is (almost) flat along the $k_{x}$ direction of the surface BZ \cite{hao11}. It was argued in later works that, when the Fermi surface becomes two-dimensional-like with copper intercalation, the SABSs for $\underline{\Delta}_{4a}$ would disappear \cite{mizushima14,fu14}. This conclusion is natural if the Fermi surface is purely cylindrical with no dispersion along $k_{z}$, because the existence of SABSs on the $xy$ surface is associated with a sign change in the pairing term upon reflection from the surface, which requires on one hand a finite dispersion along $k_{z}$ and on the other hand a pairing component that changes sign with the reversal of $k_{z}$.

However, there seems to be no reason why the Fermi surface can turn from three-dimensional to completely two-dimensional with copper intercalation. On one hand, one experiment reports corrugated cylindrical rather than completely cylindrical Fermi surface \cite{lahoud13}. On the other hand, Cu$_{x}$Bi$_{2}$Se$_{3}$ is known to be superconducting in a wide range of $x$ values \cite{hor10,kriener11}. It is natural to expect that the Fermi surface evolves continuously from spheroidal to corrugated cylindrical as $x$ increases. Finally, a completely two-dimensional Fermi surface is inconsistent with the existence of the TSSs, observed experimentally \cite{wray10,lahoud13}. Therefore, compared to completely cylindrical Fermi surface, corrugated cylindrical Fermi surfaces with different degrees of corrugation [e.g., those shown in Fig.1] are better descriptions of the actual Fermi surface. For these Fermi surfaces, the bulk gap for the $\underline{\Delta}_{4a}$ pairing has minima along the $k_{x}$ direction, which is explained above and illustrated in Fig.2. Since the conduction band has a finite dispersion along $k_{z}$, and the gap is dominated by the $c_{z}(\mathbf{k})$ term which is odd in $k_{z}$, it seems natural to expect the prevalent existence of SABSs. As shown below, this is true. We first give some analytical analysis, which is then followed by numerical results. Finally, we study the stability of the SABSs against surface and bulk nonmagnetic impurities.

\subsection{Analytical analysis for a simplified model}

To gain a qualitative understanding of the SABSs for a Fermi surface in the shape of a spheroid or a corrugated cylinder, we ignore the hexagonal warping in the Fermi surface and consider a band with the following dispersion
\begin{equation}
\xi_{\mathbf{k}}=\frac{k_{x}^{2}+k_{y}^{2}}{2m_{1}^{\ast}}-t_{z}\cos k_{z}-\mu,
\end{equation}
where $m_{1}^{\ast}$ is the effective mass in the $k_{x}k_{y}$ plane, and the dispersion along $k_{z}$ is determined by $t_{z}$. $\hbar=1$ is assumed. $m_{1}^{\ast}$ is assumed to be small so that the Fermi momenta along directions in the $k_{x}k_{y}$ plane are small, i.e. $\sqrt{2m_{1}^{\ast}\mu}$ is small ($m_{1}^{\ast}>0$ and $\mu>0$). The Fermi surface is spheroidal (corrugated cylindrical) when $|t_{z}/\mu|\ge1$ ($0<|t_{z}/\mu|<1$).

For the pairing term, it is convenient to replace the factor $E_{\mathbf{k}}$ with a constant. Eq.(17) is thus reduced to
\begin{eqnarray}
\tilde{\underline{\Delta}}_{4a}(\mathbf{k})&=&\tilde{\Delta}_{0}[-R_{1}d_{1}(\mathbf{k})\rho_{1} -(B_{0}c_{z}(\mathbf{k})+R_{2}d_{2}(\mathbf{k}))\rho_{2}     \notag \\
&&+A_{0}c_{y}(\mathbf{k})\rho_{3}]i\rho_{2},
\end{eqnarray}
where $\tilde{\Delta}_{0}$ is the pairing amplitude divided by the constant representing $E_{\mathbf{k}}$.

There are several available approaches that we can use to derive the SABSs. Here, we follow the approach of mapping the surface problem by an equivalent junction problem. Namely, we consider a junction at $z=0$ between the surfaces of two bulk samples; one is extended from $z=0$ to $\infty$, and the other  is from $z=0$ to $-\infty$. The two bulk samples are both described by Eq.(21). The problem of scattering off the surface is mapped to a sign change in the components of the pairing term odd in $k_{z}$. The $z<0$ part of the junction is described simply by Eqs.(21) and (22). The $z>0$ part of the junction is described by Eq.(21) and Eq.(22) with the sign of the term proportional to $c_{z}(\mathbf{k})$ reversed. The model on either side of the junction is a $4\times4$ model with two $2\times2$ block diagonals for the bare bands and two $2\times2$ off-diagonal blocks representing the pairing term.

To proceed, we adopt the quasiclassical approximation to the ansatz for the wave function of the SABSs, to separate the fast and slow degrees of freedom. The $4\times4$ eigenvector of the SABSs is thus taken as
\begin{equation}
\varphi(\mathbf{k}_{F},\mathbf{r})=\begin{pmatrix} u(\mathbf{r}) \\ v(\mathbf{r}) \end{pmatrix}=
e^{i\mathbf{k}_{F}\cdot\mathbf{r}}\begin{pmatrix} f(\mathbf{r}) \\ g(\mathbf{r}) \end{pmatrix}.
\end{equation}
In the same spirit, we expand the bulk band around the Fermi momentum $\mathbf{k}_{F}$ as
\begin{equation}
\xi_{\mathbf{k}}\simeq \mathbf{v}_{F}(\mathbf{k}_{F})\cdot(-i\boldsymbol{\nabla}-\mathbf{k}_{F}),
\end{equation}
where the Fermi velocity is defined as
\begin{equation}
\mathbf{v}_{F}(\mathbf{k}_{F})=\boldsymbol{\nabla}_{\mathbf{k}}\xi_{\mathbf{k}}|_{\mathbf{k}=\mathbf{k}_{F}}.
\end{equation}
We assume a perfect junction in which the translational invariance within the junction plane is preserved. The problem is thus reduced to a one-dimensional scattering problem along the $z$ direction. That is, the dependencies in the $x$ and $y$ coordinates occur only through the exponential pre-factor of Eq.(23). $f$ and $g$ depend only on $z$. Consistent with this assumption on the wave function, we replace in the pairing term the $k_{x}$ and $k_{y}$ components of the wave vectors with $k_{Fx}$ and $k_{Fy}$. The $c_{z}(\mathbf{k})$ term is then expanded to linear order of $k_{z}-k_{Fz}$.

The interface localized states are solved by imposing the following boundary conditions to the wave function
\begin{equation}
\varphi(\mathbf{k}_{F},x,y,z=0^{+})=\varphi(\mathbf{k}_{F},x,y,z=0^{-}),
\end{equation}
\begin{equation}
\varphi(\mathbf{k}_{F},x,y,z=-\infty)=\varphi(\mathbf{k}_{F},x,y,z=+\infty)=0.
\end{equation}
$0^{+}$ and $0^{-}$ are positive and negative infinitesimals. Focusing on the direction of $k_{Fy}=0$, where the minima of the superconducting gap are attained, we indeed find solutions satisfying the above boundary conditions, with energies
\begin{equation}
|E(k_{Fx},0,k_{Fz})|=|\tilde{\Delta}_{0}R_{1}d_{1}(k_{Fx},0,k_{Fz})|.
\end{equation}
From the discussions on the bulk superconducting gap in the previous section, the energy of the above bound states are well below the bulk gap in a large part of the bulk gap. Therefore, they are well-defined in-gap states.

\subsection{Numerical results for clean system}

Inspired by the existing experiments, we consider the three sets of parameters in Table I, which result separately a spheroidal Fermi surface and two corrugated cylindrical Fermi surfaces with different degrees of corrugation. Besides the shape of the Fermi surface, these parameters allow the simultaneous presence of bulk conduction band and the topological surface states (TSSs) at the Fermi level. As shown in Figure 3 are the surface spectral functions for the three parameter sets, for clean systems. The surface spectral function for a wave vector in the surface BZ are defined as summation over the imaginary part of the particle Green's function, which are obtained in terms of standard iterative Green's function method \cite{wang10,hao11,hao13}. From Fig.3, the SABSs exist for both spheroidal and corrugated cylindrical Fermi surfaces. One essential feature is the existence of a nearly flat band of Andreev bound states along $k_{x}$ at the center of the SC gap. The magnitude of the bulk gap depends sensitively on the model parameters and can be vanishingly small for the parameters with (corrugated) cylindrical Fermi surface, consistent with the previous section. A qualitative difference from previous results obtained for simplified models without the hexagonal warping terms is that \cite{hao11,sasaki11,yamakage12}, the SABSs are not exactly flat along $(k_{x},0)$, which becomes increasingly clear as the size of the (larger) gap minimum increases from Fig.1(a) to Fig.1(c) and Fig.1(e).

The surface spectral functions can be probed by ARPES \cite{wray10}. The integrated surface spectral function, the surface density of states (SDOS), can be probed by tunneling spectroscopy \cite{sasaki11,peng13,levy12}. In Fig.4, we have shown the SDOS together with the corresponding bulk density of states (BDOS). A common characteristic of the results for all three parameter sets is the appearance of a prominent zero-energy peak corresponding to the mid-gap Majorana bound states plus a continuum of low-energy states filling up the bulk superconducting gap. As is explained in Sec.IV, the minimum of the bulk gap scales linearly with $R_{1}$. We have made test calculations by increasing $R_{1}$ artificially to 2 eV and 20 eV, for the third set of parameters (`Cylindrical 2' in Table I). The minimal size of the bulk gap along $k_{x}$ increases linearly with $R_{1}$, and the BDOS becomes increasingly $U$ shaped with a flat bottom of zero DOS. The SABSs, on the other hand, persist and traverse the bulk gap for all parameters considered. The zero-energy TSSs also persist. As a result, the SDOS is still featured by the existence of in-gap states with a peak at or close to zero energy.

\begin{figure}\label{fig3} \centering
\includegraphics[width=8.5cm,height=12.5cm]{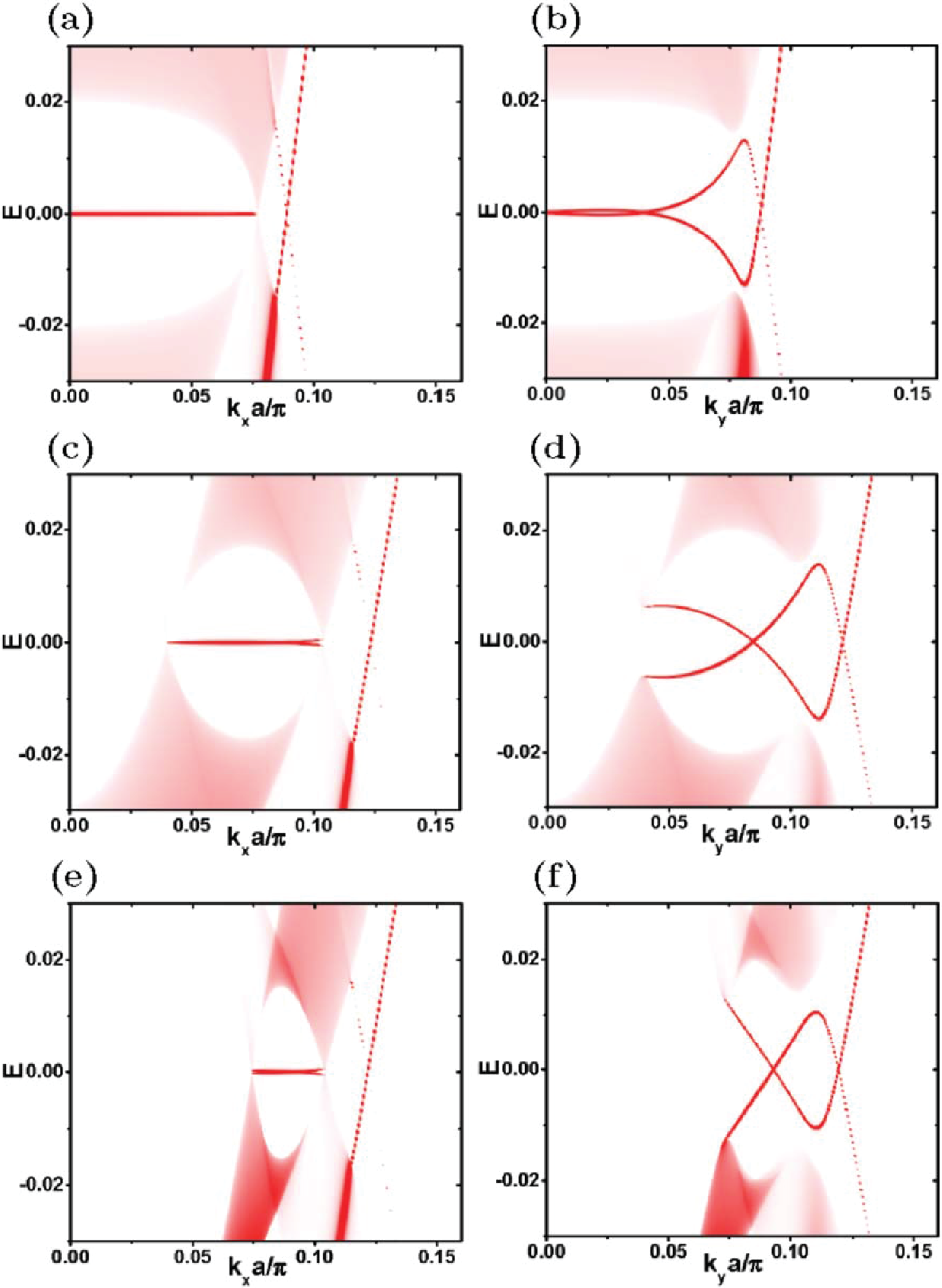} \\
\caption{(Color online) Surface spectral functions of the system with $\underline{\Delta}_{4a}$ pairing, for clean system. (a) and (b) correspond to the first (`Spheroidal') set of parameters in Table I. (c) and (d) correspond to the second (`Cylindrical 1') set of parameters in Table I. (e) and (f) correspond to the third (`Cylindrical 2') set of parameters in Table I. $\Delta_{a}=0.02$ eV. The energy $E$ is in unit of electron volts.}
\end{figure}

\begin{figure}\label{fig4} \centering
\includegraphics[width=7.0cm,height=11.5cm]{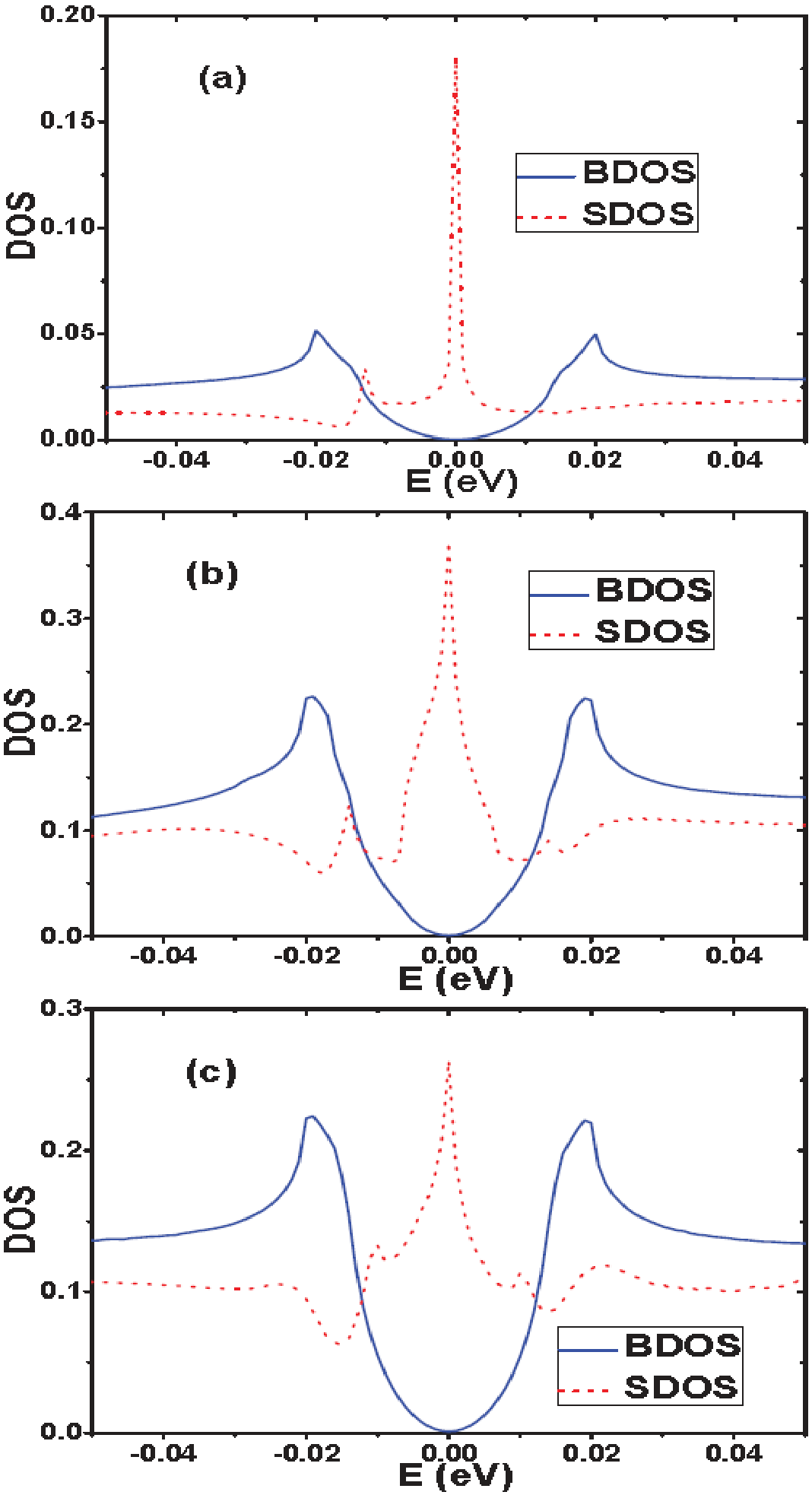} \\
\caption{(Color online) The density of states in the bulk (BDOS) and on the surface (SDOS), for clean system. $\Delta_{a}=0.02$ eV. (a), (b), and (c) are separately for the first (`Spheroidal'), second (`Cylindrical 1'), and third (`Cylindrical 2') sets of parameters in Table I.}
\end{figure}

\subsection{Stability against nonmagnetic impurities}

Having verified the presence of SABSs and seen their peculiar dispersions in clean systems, we proceed to test their stability against various imperfections. From a practical point of view relevant to Cu$_{x}$Bi$_{2}$Se$_{3}$, it is plausible to focus on the effect of short-range nonmagnetic impurities. We study the effect of the nonmagnetic impurities at two levels, impurities uniformly distributed in the whole sample and separate point-like impurities situating on the surface.

Firstly, we consider the effect of impurities distributed uniformly throughout the whole sample. We first obtain the self-energy correction to the bulk Green's functions in terms of the self-consistent $T$-matrix approximation \cite{nagai15}. We consider the simplest case of uniformly distributed short-range nonmagnetic impurities, $V(\mathbf{r})=V_{0}\delta(\mathbf{r}-\mathbf{r}_{0})$. In this case, the self-energy is $\mathbf{k}$-independent and is determined by a set of three self-consistent equations: (1) $G(\mathbf{k},\omega)=[\omega+i\eta-H(\mathbf{k})-\Sigma(\omega)]^{-1}$, (2) $\Sigma(\omega)=n_{imp}[T(\omega)-\tilde{V}]$, and (3) $T(\omega)=[I_{8}-\frac{\tilde{V}}{N}\sum_{\mathbf{k}}G(\mathbf{k},\omega)]^{-1}\tilde{V}$. Here, $n_{imp}$ is the concentration of the nonmagnetic impurity, $\tilde{V}=V_{0}\tau_{3}\otimes\sigma_{0}\otimes s_{0}$, $\mathbf{k}$ denotes a wave vector in the 3D BZ, and $N$ is the number of wave vectors in the 3D BZ. After obtaining the self-energy $\Sigma(\omega)$ from the above self-consistency loop, we add it as an energy correction to $H(\mathbf{k})$ and obtain the surface Green's function in terms of the iterative Green's function method \cite{hao11,hao13}. The resulting surface Green's function is then the proper Green's function for the surface layer in the presence of short-range nonmagnetic impurities uniformly distributed in the bulk. The bulk density of states (BDOS) and surface density of states (SDOS) obtained by this method are as shown in Figure 5. Three concentrations of the impurities ($n_{imp}=0.001$, $0.01$ and $0.02$) are considered. As the bulk superconducting gap of the cases with corrugated cylindrical (spheroidal) Fermi surface is filled up for $n_{imp}=0.01$ ($n_{imp}=0.02$), the fine structures in the SDOS beyond $E=0$ disappear, but the zero-energy surface states are still quite robust and manifest as a single zero-energy peak in the SDOS.

\begin{figure}\label{fig5} \centering
\includegraphics[width=8.5cm,height=10.5cm]{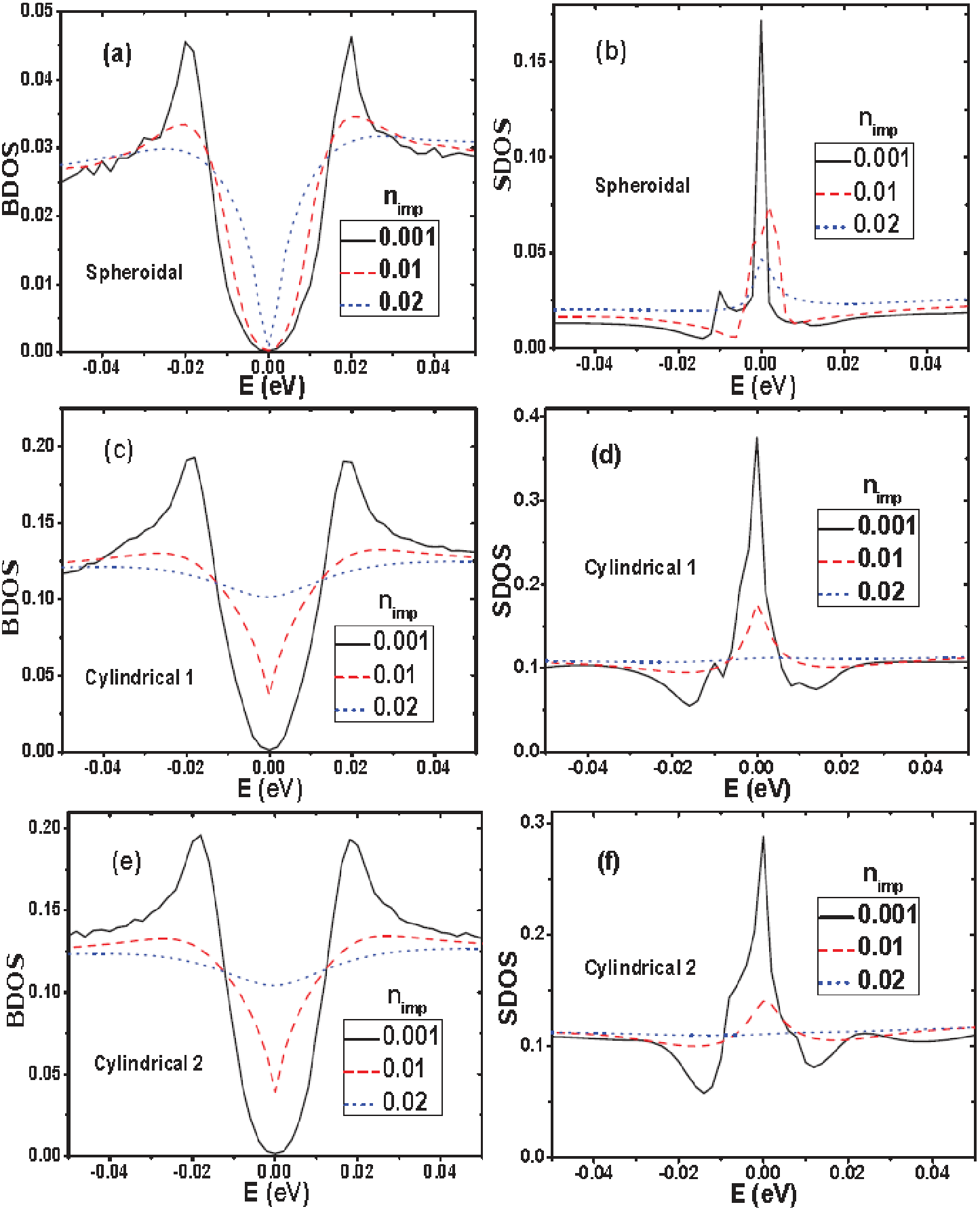} \\
\caption{(Color online) Bulk density of states (BDOS) and surface density of states (SDOS) in the presence of short-range nonmagnetic impurities uniformly distributed in the bulk. $V_{0}=10$ eV. Three impurity concentrations are considered, including $n_{imp}=0.001$, $0.01$, and $0.02$. (a) and (b) correspond to the first (`Spheroidal') set of parameters in Table I. (c) and (d) correspond to the second (`Cylindrical 1') set of parameters in Table I. (e) and (f) correspond to the third (`Cylindrical 2') set of parameters in Table I. $\mu=0.32$ eV and $\Delta_{a}=0.02$ eV are used for all calculations.}
\end{figure}

Secondly, we consider the effect of individual point-like impurities on the surface of an otherwise clean sample. This is achieved by keeping the bulk of the material clean, and adding impurities only to the surface layer in a manner that different impurities are far away from each other. For this case, we study the changes in the surface Green's functions for the clean system induced by the surface impurities. The effect of the impurities is taken in to account in terms of the $T$-matrix approximation \cite{wang10,balatsky06}. For a single short-range nonmagnetic impurity, $V(\mathbf{r})=V_{0}\delta(\mathbf{r}-\mathbf{r}_{0})$, the $T$-matrix is $\mathbf{k}$-independent
\begin{equation}
T(\omega)=[I_{8}-\frac{\tilde{V}}{N_{xy}}\sum_{\mathbf{k}}G_{0}(\mathbf{k},\omega)]^{-1}\tilde{V},
\end{equation}
where $N_{xy}$ is the number of unit cells (wave vectors) in the $xy$ plane (surface BZ), $G_{0}(\mathbf{k},\omega)$ is the retarded surface Green's function obtained in terms of the iterative Green's function method for a clean system \cite{hao11,hao13}. $\tilde{V}=V_{0}\tau_{3}\otimes\sigma_{0}\otimes s_{0}$ is the impurity potential in the Nambu space. In terms of the unperturbed Green's function $G_{0}(\mathbf{k},\omega)$ and the $T$-matrix, the perturbed Green's function is obtained in term of the $T$-matrix approximation as
\begin{equation}
G(\mathbf{r},\mathbf{r}',\omega)=G_{0}(\mathbf{r}-\mathbf{r}',\omega)+G_{0}(\mathbf{r}-\mathbf{r}_{0},\omega)T(\omega)G_{0}(\mathbf{r}_{0}-\mathbf{r}',\omega),
\end{equation}
where $G_{0}(\mathbf{r},\omega)$ is the Fourier transformation of $G_{0}(\mathbf{k},\omega)$. The local Green's function at $\mathbf{r}$ under the influence of the nonmagnetic impurity at $\mathbf{r}_{0}$ is defined as $G(\mathbf{r},\mathbf{r},\omega)$. We consider the effect of a strong unitary impurity and take $V_{0}=1000$ eV. The results of the local density of states (SLDOS) at the impurity site and its six nearest-neighbor sites for the three sets of parameters in Table I are shown in Fig. 6. The SLDOS on the nearest-neighbor site $\mathbf{r}-\mathbf{r}_{0}=\boldsymbol{\delta}_{\alpha}$ is equal to the SLDOS on the nearest-neighbor site $\mathbf{r}-\mathbf{r}_{0}=-\boldsymbol{\delta}_{\alpha}$ ($\alpha$=1,2,3), so only results for three of the six nearest-neighbor sites are shown. Albeit quantitative differences from the SDOS on Fig.4 and Fig.5, the in-gap surface states persist. On the other hand, the SLDOS for $\mathbf{r}-\mathbf{r}_{0}=\boldsymbol{\delta}_{1}$ and $\mathbf{r}-\mathbf{r}_{0}=\boldsymbol{\delta}_{2}$ are identical but are different from the SLDOS for $\mathbf{r}-\mathbf{r}_{0}=\boldsymbol{\delta}_{3}$, which is consistent with the two-fold anisotropy of the superconducting pairing between the $x$ and $y$ directions. It is also noted that the zero-energy surface states are also very robust at and near the impurity site for the three types of Fermi surfaces (see Fig. 6).

\begin{figure}\label{fig6} \centering
\includegraphics[width=7.0cm,height=11.5cm]{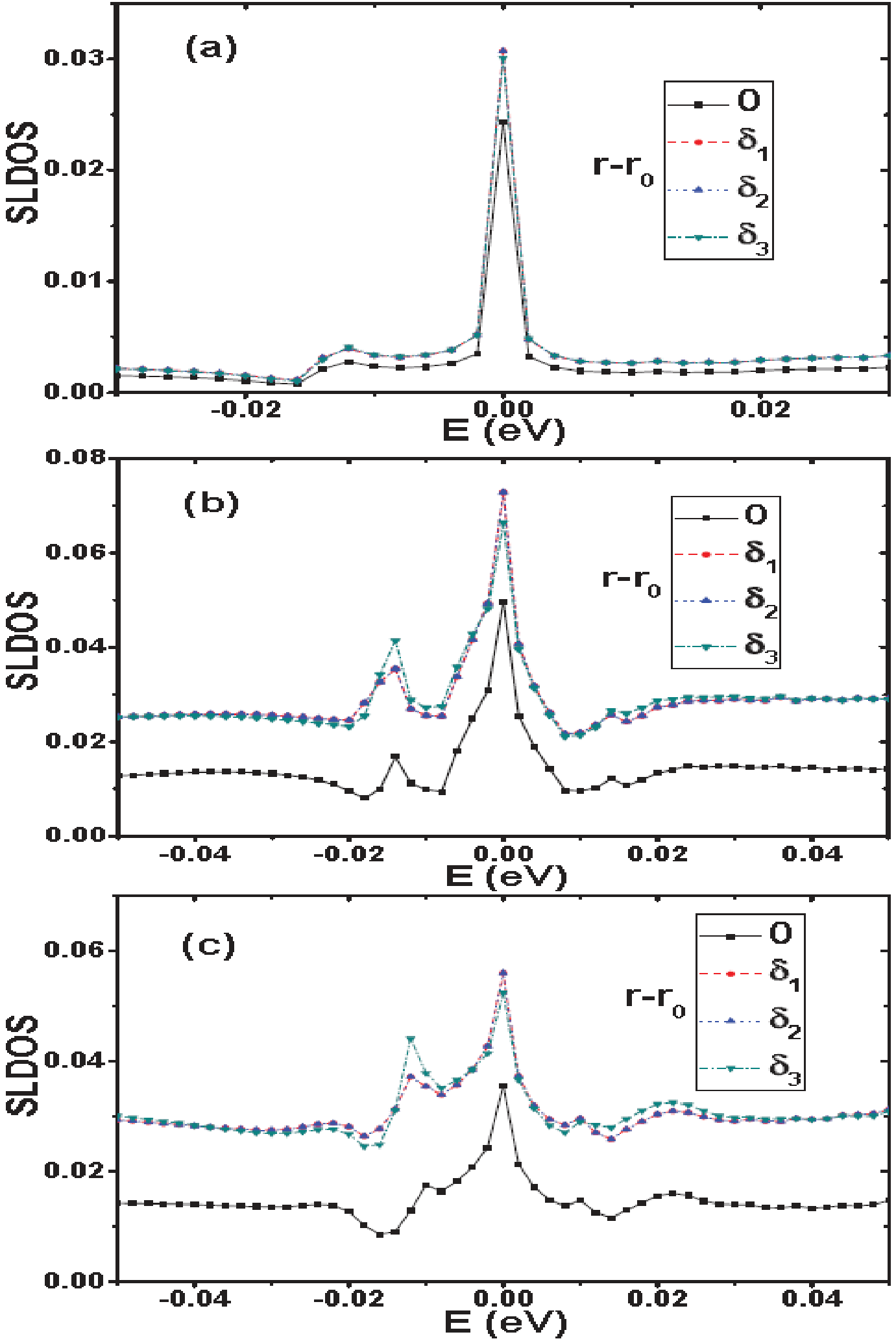} \\
\caption{(Color online) The surface local density of states (SLDOS) at the impurity site ($\mathbf{r}-\mathbf{r}_{0}=\mathbf{0}$) and in the nearest-neighboring sites ($\mathbf{r}-\mathbf{r}_{0}=\boldsymbol{\delta}_{\alpha}$, $\alpha$=1,2,3) of a strong ($V_{0}=1000$ eV) point-like nonmagnetic impurity. $\Delta_{a}=0.02$ eV. (a), (b), and (c) are separately for the first (`Spheroidal'), second (`Cylindrical 1'), and third (`Cylindrical 2') sets of parameters in Table I.}
\end{figure}

\section{conclusion}

In conclusion, extensive analyses for the bulk and surface spectra have been made for the $\underline{\Delta}_{4a}(\mathbf{k})$ nematic pairing, suggested to be the correct SC pairing for the Cu$_{x}$Bi$_{2}$Se$_{3}$ compound based on recent Knight shift and field-angle dependent specific heat measurements\cite{matano16,yonezawa17}. The purpose of the present work is to explore the consequences deduced from this type of pairing, taking into account the evolution of the Fermi surface from spheroidal to corrugated cylindrical \cite{lahoud13}. We show that Cu$_{x}$Bi$_{2}$Se$_{3}$ with $\underline{\Delta}_{4a}(\mathbf{k})$ pairing should be a topological superconductor with topological surface states and the surface Andreev bound states (SABSs), even if the Fermi surface has changed from spheroidal \cite{hao11,sasaki11,yamakage12} to the corrugated cylindrical case studied in the present work. The bulk SC spectrum, while fully gapped, show prominent twofold anisotropy with vanishingly small gap minima along $k_{x}$. One of the essential features of the SABSs is the exhibition of the zero-energy Majorana bound states regardless the shape of the Fermi surface. The SABSs are shown to be robust against short-range bulk and surface nonmagnetic impurities. This is consistent with recent works which show that the surface states of class DIII topological superconductor are stable against weak disorder and interaction \cite{foster14,xie15}. On the other hand, there are experimental controversies on the existence of SABSs \cite{sasaki11,koren11,kirzhner12,levy12,peng13}. As to whether the SABSs could be observed experimentally may depend critically on the condition of the sample surfaces. For instance, the excessive magnetic Cu$^{2+}$ ($3d^9$) ions or Cu ($3d^{10}4S^{1}$) atoms on the surfaces could very much suppress the SABSs. For clean and perfect surface, on the other hand, the SABSs should be detectable. The present pairing model, an odd-parity spin-triplet pairing, yields a very anisotropic bulk SC gap while the STM experiment detected an almost isotropic s-wave like bulk gap\cite{levy12}. It appears that the present nematic pairing model, which has successfully explained recent Knight shift and field-angle dependent specific heat measurements, are apparently having difficulty to account for other experimental measurements like the SC density of states \cite{levy12}. Therefore it is necessary to develop a revised pairing model which includes the essential physics as discussed here and also being able to explain other experiments. This will definitely constitute a challenging topic for future study.

\begin{acknowledgments}
L.H. thanks Ting-Kuo Lee and Sungkit Yip for many helpful discussions. This work was supported in part by the Robert A. Welch Foundation under Grant No. E-1146 and the Texas Center for Superconductivity at the University of Houston. L.H. is also supported by NSFC.11204035 and would also like to acknowledge the support from the China Scholarship Council. Part of the calculations were performed in the Center for Advanced Computing and Data Systems in University of Houston.
\end{acknowledgments}\index{}



\begin{references}

\bibitem{hor10} Y. S. Hor, A. J. Williams, J. G. Checkelsky, P. Roushan,
J. Seo, Q. Xu, H. W. Zandbergen, A. Yazdani, N. P. Ong,
and R. J. Cava, Phys. Rev. Lett. \textbf{104}, 057001 (2010).


\bibitem{wray10} L. Andrew Wray, Su-Yang Xu, Yuqi Xia, Yew San Hor,
Dong Qian, Alexei V. Fedorov, Hsin Lin, Arun Bansil,
Robert J. Cava and M. Zahid Hasan, Nature Phys. \textbf{1762}
(2010).


\bibitem{sasaki11} Satoshi Sasaki, M. Kriener, Kouji Segawa, Keiji Yada,
Yukio Tanaka, Masatoshi Sato, and Yoichi Ando, Phys.
Rev. Lett. \textbf{107}, 217001 (2011).

\bibitem{koren11} G. Koren, T. Kirzhner, E. Lahoud, K. B. Chashka, and
A. Kanigel, Phys. Rev. B \textbf{84}, 224521 (2011).

\bibitem{kirzhner12} T. Kirzhner, E. Lahoud, K. B. Chaska, Z. Salman, and
A. Kanigel, Phys. Rev. B \textbf{86}, 064517 (2012).

\bibitem{kriener11} M. Kriener, Kouji Segawa, Zhi Ren, Satoshi Sasaki, and
Yoichi Ando, Phys. Rev. Lett. \textbf{106}, 127004 (2011).

\bibitem{das11} Pradip Das, Yusuke Suzuki, Masashi Tachiki, and Kazuo
Kadowaki, Phys. Rev. B \textbf{83}, 220513(R) (2011).

\bibitem{bay12} T.V. Bay, T. Naka, Y. K. Huang, H. Luigjes, M. S.
Golden, and A. de Visser, Phys. Rev. Lett. \textbf{108}, 057001
(2012).


\bibitem{levy12} N. Levy, T. Zhang, J. Ha, F. Sharifi, A. A. Talin, Y. Kuk, and
J. A. Stroscio, Phys. Rev. Lett. \textbf{110}, 117001 (2013).


\bibitem{peng13} Haibing Peng, Debtanu De, Bing Lv, Fengyan Wei, and Ching-Wu Chu, Phys. Rev. B \textbf{88}, 024515 (2013).


\bibitem{matano16} K. Matano, M. Kriener, K. Segawa, Y. Ando, and Guo-qing Zheng, Nat. Phys. \textbf{12}, 852 (2016).

\bibitem{fu16} Liang Fu, Nat. Phys. \textbf{12}, 822 (2016).

\bibitem{zhao14} Lukas Zhao, Haiming Deng, Inna Korzhovska,	Milan Begliarbekov,	Zhiyi Chen,	Erick Andrade, Ethan Rosenthal,	Abhay Pasupathy, Vadim Oganesyan, Lia Krusin-Elbaum, Nature Commun. \textbf{6}, 8279 (2014).


\bibitem{pan16} Y. Pan, A. M. Nikitin, G. K. Araizi, Y. K. Huang, Y. Matsushita, T. Naka, and A. de Visser, Sci. Rep. \textbf{6}, 28632 (2016).


\bibitem{du17} Guan Du, YuFeng Li, J. Schneeloch, R. D. Zhong, GenDa Gu, Huan Yang, Hai Lin, and Hai-Hu Wen, Sci. China-Phys. Mech. Astron. \textbf{60}, 037411 (2017)

\bibitem{qiu15} Y. Qiu, K. N. Sanders, J. Dai, J. E. Medvedeva, W. Wu, P. Ghaemi, T. Vojta, and Y. S. Hor, arXiv:1512.03519.


\bibitem{asaba17} Tomoya Asaba, B. J. Lawson, Colin Tinsman, Lu Chen, Paul Corbae, Gang Li, Y. Qiu, Y. S. Hor, Liang Fu, and Lu Li, Phys. Rev. X \textbf{7}, 011009 (2017).


\bibitem{lawson16} B. J. Lawson, Paul Corbae, Gang Li, Fan Yu, Tomoya Asaba, Colin Tinsman, Y. Qiu, J. E. Medvedeva, Y. S. Hor, and Lu Li, Phys. Rev. B \textbf{94}, 041114(R) (2016).


\bibitem{wang16} Zhiwei Wang, A. A. Taskin, Tobias Fr\"{o}lich, Markus Braden, and Yoichi Ando, Chem. Mater. \textbf{28}, 779 (2016).




\bibitem{yonezawa17} Shingo Yonezawa, Kengo Tajiri, Suguru Nakata, Yuki Nagai, ZhiweiWang, Kouji Segawa,
Yoichi Ando, and Yoshiteru Maeno, Nature Phys. \textbf{13}, 123 (2017).


\bibitem{fu10} Liang Fu and Erez Berg, Phys. Rev. Lett. \textbf{105}, 097001
(2010).

\bibitem{fu14} Liang Fu, Phys. Rev. B \textbf{90}, 100509 (2014).



\bibitem{yip13} S.-K. Yip, Phys. Rev. B \textbf{87}, 104505 (2013).


\bibitem{hao14} Lei Hao and Ting-Kuo Lee, J. Phys.: Condens. Matter \textbf{27}, 105701 (2015); \emph{ibid} arXiv:1407.3329v2.



\bibitem{hao11} Lei Hao and T. K. Lee, Phys. Rev. B \textbf{83}, 134516 (2011).


\bibitem{yamakage12} Ai Yamakage, Keiji Yada, Masatoshi Sato, and Yukio Tanaka, Phys. Rev. B \textbf{85}, 180509(R) (2012).


\bibitem{hao15} Lei Hao and Jun Wang, J. Phys.: Condens. Matter \textbf{27}, 255701 (2015).



\bibitem{lahoud13} E. Lahoud, E. Maniv, M. S. Petrushevsky, M. Naamneh, A. Ribak, S. Wiedmann, L. Petaccia, Z. Salman, K. B. Chashka, Y. Dagan, and A. Kanigel, Phys. Rev. B \textbf{88}, 195107 (2013).


\bibitem{mizushima14} Takeshi Mizushima, Ai Yamakage, Masatoshi Sato, and Yukio Tanaka, Phys. Rev. B \textbf{90}, 184516 (2014).



\bibitem{liu15} Zhongheng Liu, Xiong Yao, Jifeng Shao, Ming Zuo, Li Pi, Shun Tan, Changjin Zhang,
and Yuheng Zhang, J. Am. Chem. Soc. \textbf{137}, 10512 (2015).


\bibitem{nagai15} Yuki Nagai, Phys. Rev. B \textbf{91}, 060502(R) (2015).



\bibitem{liu10} Chao-Xing Liu, Xiao-Liang Qi, Haijun Zhang,
    Xi Dai, Zhong Fang, and Shou-Cheng Zhang, Phys. Rev. B \textbf{82},
    045122 (2010).


\bibitem{fu09} Liang Fu, Phys. Rev. Lett. \textbf{103}, 266801
    (2009).


\bibitem{zhang09} Haijun Zhang, Chao-Xing Liu, Xiao-Liang Qi,
    Xi Dai, Zhong Fang, and Shou-Cheng Zhang, Nature Phys.
\textbf{5}, 438 (2009).


\bibitem{wang10} Qiang-Hua Wang, Da Wang, and Fu-Chun Zhang,
    Phys. Rev. B \textbf{81}, 035104 (2010).


\bibitem{fu07} Liang Fu and C. L. Kane, Phys. Rev. B \textbf{76}, 045302 (2007).



\bibitem{hao13} Lei Hao, Gui-Ling Wang, Ting-Kuo Lee, Jun Wang, Wei-Feng Tsai, and Yong-Hong Yang, Phys. Rev. B \textbf{89}, 214505 (2014).



\bibitem{acparameters}  P. Larson, V. A. Greanya, W. C. Tonjes, Rong Liu, S. D. Mahanti, C. G. Olson,
     Phys. Rev. B \textbf{65}, 085108 (2002).



\bibitem{yip14} Sungkit Yip, Annu. Rev. Cond. Matter Physics \textbf{5}, 15 (2014).


\bibitem{yip16} Sungkit Yip, arXiv:1609.04152.


\bibitem{zocher13} Bj\"{o}rn Zocher and Bernd Rosenow, Phys. Rev. B \textbf{87}, 155138 (2013).


\bibitem{hashimoto13} Tatsuki Hashimoto, Keiji Yada, Ai Yamakage, Masatoshi Sato, and Yukio Tanaka, J. Phys. Soc. Jpn. \textbf{82}, 044704 (2013).


\bibitem{nagai14} Yuki Nagai, Hiroki Nakamura, and Masahiko Machida, J. Phys. Soc. Jpn. \textbf{83}, 053705 (2014).


\bibitem{takami14} Shota Takami, Keiji Yada, Ai Yamakage, Masatoshi Sato, and Yukio Tanaka, J. Phys. Soc. Jpn. \textbf{83}, 064705 (2014).



\bibitem{fu15} Liang Fu, Phys. Rev. Lett. \textbf{115}, 026401 (2015).


\bibitem{kozii15} Vladyslav Kozii and Liang Fu, Phys. Rev. Lett. \textbf{115}, 207002 (2015).


\bibitem{venderbos16a} J\"{o}rn W. F. Venderbos, Vladyslav Kozii, and Liang Fu, Phys. Rev. B \textbf{94}, 094522 (2016).


\bibitem{venderbos16b} J\"{o}rn W. F. Venderbos, Vladyslav Kozii, and Liang Fu, Phys. Rev. B \textbf{94}, 180504(R) (2016).



\bibitem{note1} Qualitatively the same effective pairing can be obtained in terms of a simplified method of deriving the low-energy effective model \cite{hao14}. The only change is to substitute $\mu+M(\mathbf{k})-\epsilon(\mathbf{k})$ for $E_{\mathbf{k}}$ in the denominator of Eq.(17). The simplified method thus gives the same picture about the nature of the pairing, at the expense of less quantitative accuracy \cite{hao14}.



\bibitem{balatsky06} A. V. Balatsky, I. Vekhter, and Jian-Xin Zhu, Rev. Mod. Phys. \textbf{78}, 373 (2006).



\bibitem{foster14} Matthew S. Foster, Hong-Yi Xie, and Yang-Zhi Chou, Phys. Rev. B \textbf{89}, 155140 (2014).


\bibitem{xie15} Hong-Yi Xie, Yang-Zhi Chou, and Matthew S. Foster, Phys. Rev. B \textbf{91}, 024203 (2015).


\end{references}
\end{document}